\documentclass[]{llncs}
\usepackage{float}
\usepackage{caption}
\usepackage{subcaption}
\usepackage{url}
\usepackage{paralist}
\usepackage{todonotes}
\usepackage[pdfpagelabels,colorlinks,allcolors=blue]{hyperref}
\usepackage{xspace}
\usepackage{algorithm}
\usepackage{algpseudocode}
\usepackage{amsfonts}
\usepackage{amsmath}

\newcommand{\tablesize}{0.22}

\usepackage{graphicx}
\graphicspath{{Figures/}}
\begin{document}

\title{Spherical Graph Drawing by Multi-dimensional Scaling}
\author{Jacob Miller \and Vahan Huroyan \and Stephen Kobourov \\
        \small{\url{jacobmiller1@arizona.edu} \and \url{vahanhuroyan@math.arizona.edu} \and \url{kobourov@cs.arizona.edu}}}

\institute{Department of Computer Science, University of Arizona}

%\author{}
%
%\authorrunning{}
%\institute{} 

%
\maketitle              % typeset the header of the contribution

\begin{abstract}
    We describe an efficient and scalable spherical graph embedding method.  
    %on spheres 
    %such that the graph theoretical distances between all pairs of graph's vertices match with the geodesic distances between the embedded ones on sphere.
    %{\color{red}such that the embedded geodesic distances closely match the graph-theoretic distances between vertices.}
    The method uses a generalization of the Euclidean stress function for  Multi-Dimensional Scaling adapted to spherical space, where geodesic pairwise distances are employed instead of Euclidean  distances. The resulting spherical stress function is optimized by means of stochastic gradient descent. 
    %We numerically demonstrate that the algorithm is able to handle large graphs.
    Quantitative and qualitative evaluations demonstrate the scalability and effectiveness of the proposed method.
    %{\color{red}We numerically and visually investigate the quality of layouts produced by our method and numerically demonstrate the algorithm's scalability.}     
    We also show that some graph families can be embedded with lower distortion on the sphere, than in Euclidean and hyperbolic spaces. 
    %Finally, we consider the problem of  scaling problem seen in non-Euclidean spaces}. 

\end{abstract}

\begin{figure}
\vspace{-.8cm}
    \centering
    \includegraphics[width=0.32\textwidth]{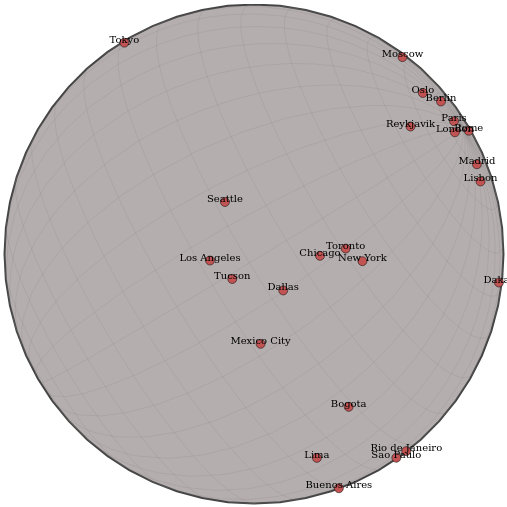}
    \includegraphics[width=0.32\textwidth]{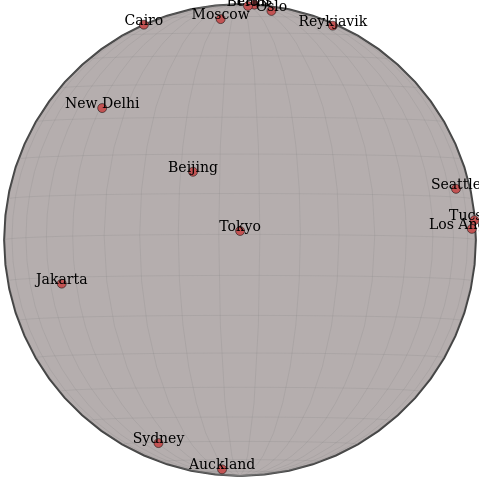}
    \includegraphics[width=0.32\textwidth]{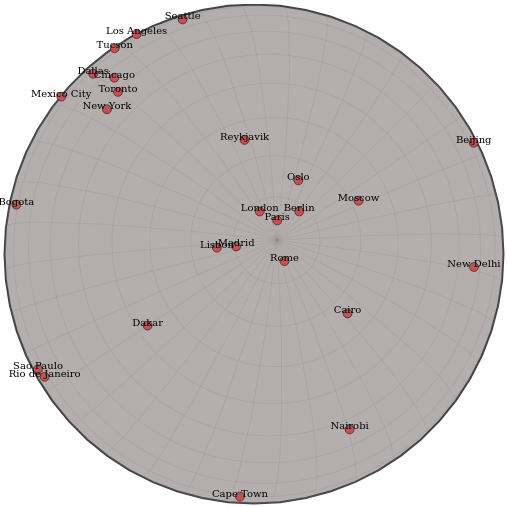}
    \caption{Applying spherical MDS to embed 30 cities from around the Earth (given pairwise distances between the cities). The spherical MDS recovers the underlying geometry.}
    \label{fig:teaser}
\end{figure}

\vspace{-.8cm}
\section{Introduction}
Node-link diagrams are typically created by embedding the vertices and edges of a given graph in the Euclidean plane and different embedding spaces are rarely considered. 
%However, not all graphs have an ideal representation in the Euclidean plane: for example 3-dimensional polytopes or triangulations of the surface of a 3-dimensional object can be better represented on the sphere, while trees and special lattices are well-suited to hyperbolic spaces. 
Multi-Dimensional Scaling (MDS), realized via stress minimization or stress majorization, is one of the standard approaches to embedding graphs in Euclidean space. 
The idea behind MDS is to place the vertices of the graph in Euclidean space so that the distances between them are as close as possible 
% as 
to
the given graph theoretic distances.
Due to the nature of Euclidean geometry, this cannot always be done without some distortion (e.g., while $K_3$ naturally lives in 2D, $K_4$ does not, as there are no four equidistant points in the Euclidean plane).
Moreover, some graphs ``live" naturally in  manifolds other than the Euclidean plane. 
For example 3-dimensional polytopes, or triangulations of 3-dimensional objects can be better represented on the sphere, while trees and special lattices are well-suited to hyperbolic spaces.  
%For example, skeletons of three dimensional polytopes such as the tetrahedron and the cube can be better visualized on spheres. 
%The Hierarchical graphs, including trees, where all edges have uniform lengths and the vertices are uniformly distributed in the space can be better visualized in hyperbolic space.

%The closeness idea is later described in this paper by an objective function.

When visualizing graphs in  Euclidean space, common techniques include adapting off-the-shelf dimensionality reduction algorithms to the graph setting. Such algorithms include the Multi-Dimensional Scaling (MDS)~\cite{shepard1962analysis}, Principal Component Analysis (PCA)~\cite{frey1978principal}, t-distributed Stochastic Neighbor Embedding (t-SNE)~\cite{van2008visualizing}, and Uniform Manifold Approximation Projection (UMAP)~\cite{mcionnes2018umap}.
The popularity of graph visualisation, and the fact that some of the underlying datasets are easier to embed in non-Euclidean spaces, led to some visualization techniques for spherical geometry~\cite{elbaz2005texture,perry2020drawing} and hyperbolic geometry~\cite{krioukov2010hyperbolic,miller2022,sala2018representation}. 
%In particular, spherical MDS aims to place vertices on the sphere so that the geodesic distances between pairs of vertices is as close as possible to the underlying graph distances, but the underlying optimization is computationally expensive.
%With this in mind, we propose a fast and accurate method for visualizing graphs on the sphere, where the spherical Multi-Dimensional Scaling (MDS) stress function is optimized by stochastic gradient descent (SGD).
Most of the existing non-Euclidean graph visualization approaches, however, either lack in accuracy or do not scale to larger graphs. 

With this in mind, we propose and analyze a stochastic gradient descent algorithm for spherical MDS.
Specifically, we present a scalable technique to compute graph layout directly on the sphere, adapting previous work for general datasets~\cite{elbaz2005texture} and applying stochastic gradient descent~\cite{robbins1951stochastic,DBLP:journals/tvcg/ZhengPG19}. We  provide an evaluation of the technique by comparing its speed and faithfulness to the exact gradient descent approach. We also investigate differences in graph layouts between the consistent geometries (Euclidean, spherical, hyperbolic) by first showing that {\em dilation or resizing} has a large effect on layouts in spherical and hyperbolic geometry, and second by showing some structures can be better
represented in one geometry than the other two. All sourcecode, datasets and  experiments, as well a web based visualization tool are available on GitHub: \url{https://github.com/Mickey253/spherical-mds}.

Note that the proposed method is not restricted to graphs, but is applicable to any dataset  specifying a set of objects and pairwise distances between them. %Moreover, ts input is a distance matrix and one can create such a matrix for a high dimensional dataset and embed it on a sphere.

% Outline
% GD paper: 
% \begin{itemize}
%     \item Spherical MDS
%     \item Spherical SGD
%     \item Arguing that graphs that naturally live in S can be embedded with lower distortion: use subvidived platonic solids or triangular meshes and compare distortion in E/S/H
%     \item Comparing SGD in E/S/H
%     \item Scaling parameter: not needed in E, definitely plays a big role in S/H
%     \item Some experiments before arriving at $\pi$
% \end{itemize}

\section{Background and Related Work}
We review related work in non-Euclidean geometry and graph layout methods.
%in non-Euclidean spaces.

\subsection{Multi-dimensional Scaling}

Using graph-theoretic distances to determine a graph layout dates back to the Kamada-Kawai algorithm~\cite{kamada1989algorithm}. A more general embedding approach from a given set of distances is the multi-dimensional scaling (MDS)~\cite{shepard1962analysis} which has extensively been applied to graph layout; see~\cite{DBLP:journals/tvcg/GansnerHN13,DBLP:conf/gd/GansnerKN04,DBLP:journals/tvcg/ZhengPG19}. Both the Kamda-Kawai and (metric) MDS algorithms aim to minimize the \textit{stress} function, which is the sum of residual squares between the given and the embedded distances of each pair of datapoints. Formally, given a graph $G = (V, E)$ with the graph theoretic distances between its $n$ vertices $\left(d_{ij}\right)_{i, j = 1}^{n, n}$, where the vertices are labeled $1, 2, \dots, n$ MDS aims to embed the graph in $\mathbb{R}^d$ by minimizing the following \textit{stress} function to find the locations for its vertices:
\begin{equation}
    \label{eq:graph_MDS}
    \sigma(X) = \sum_{i<j} w_{ij}(||X_i - X_j || - d_{ij})^2.
\end{equation}
The resulting solution of 
$X_1, X_2, \dots, X_n \in \mathbb{R}^d$ 
represents the coordinates of the embedded graph vertices.

%MDS  dates back to the 1960s when Shepard~\cite{shepard1962analysis} and  Kruskal~\cite{kruskal1964multidimensional} studied the non-metric setting. Non-metric MDS aims to recover the structure of a graph or a dataset from measures of similarity, rather than relying on the exact distances. Another variant of the MDS is the classical MDS. The classical MDS  takes a matrix of dissimilarities between pairs of objects as an input and finds an embedding that minimizes a loss function called \textit{strain}.

Various forms of MDS have been analyzed. Metric MDS was first studied by Shepard~\cite{shepard1962analysis} (see equation in~\eqref{eq:gen_stress}), and the related non-metric MDS by Kruskal~\cite{kruskal1964multidimensional}. 
Classical MDS is similar but uses an objective function called \textit{strain}.

The classical MDS has a closed form solution while the metric MDS and non-metric MDS rely on solving an optimization problems to minimize the corresponding stress functions.
Many approaches have been proposed to solve (metric) MDS including stress majorization~\cite{gansner2004graph} and (stochastic) gradient descent~\cite{bottou2010large}.

When used for the purposes of visualization, the embedding space for MDS is almost always 2 dimensional Euclidean, as that is the space of a flat sheet of paper, or the flat screen of a computer monitor. The natural measure of distance is then the Euclidean norm.

In this work we will focus on metric MDS, defined in \eqref{eq:graph_MDS} but instead of embedding the graph in Euclidean space, we embed it directly on the sphere. 
The MDS approach has already been applied to embed graphs on spherical~\cite{elbaz2005texture} and hyperbolic~\cite{miller2022} spaces. 
Our contribution is to solve the proposed optimization problem faster and be able to handle larger graphs, address the dilation/resizing problem, as well as analyze the approach on wider range of graphs and provide a working and easy to use implementation.

\subsection{Non-Euclidean Geometry}
%In this section we briefly go over the concepts of non-euclidean geometry that we need in later sections. 
Non-Euclidean geometries are a special case of Riemannian geometries, which are spaces that are locally ``smooth": one can define an inner product on the tangent space at each point. 
Spherical and hyperbolic non-Euclidean geometries are similar to Euclidean geometry, except for one axiom. 

Euclid's \textit{Elements} specify five axioms/postulates upon which all true statements about geometry should be proved. 
The fifth axiom is significantly more involved than the first four and mathematicians attempted for centuries to prove it using only the first four. In 1892 Lobachevsky and Bolyai independently discovered and published their formulation of hyperbolic geometry by inverting an equivalent statement to Euclid's fifth axiom, Playfair's axiom: \textit{In a plane, given a line and a point not on it, at most one line parallel to the given line can be drawn through the point}.
Replace ``\textit{at most one line}" with ``\textit{at least two distinct lines}" to get hyperbolic geometry. 
Replace ``\textit{at most one line}" with ``\textit{there does not exist a line}" to arrive at spherical geometry.

% Many seemingly fundamental properties of geometry are actually unique to Euclidean geometry. 
% Properties such as the Pythagorean theorem, the law of cosines, and even the fact that there is no upper bound to the area of a triangle are all emergent behavior from the lack of curvature in Euclidean space. 
% One can reduce the definition of curvature to the behavior of parallel lines. In Euclidean space, where parallel lines are infinitely parallel, the curvature is 0. There are two other options, each corresponding to a consistent geometry. 
% If parallel lines eventually converge, we say the space is positively curved and spherical. If parallel lines eventually diverge, we say the space is negatively curved and hyperbolic.

\begin{figure}[t]
    \centering
    \includegraphics[width=0.32\textwidth]{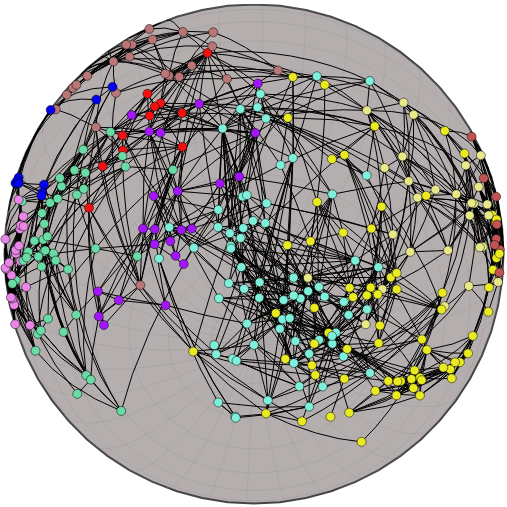}
    \includegraphics[width=0.32\textwidth]{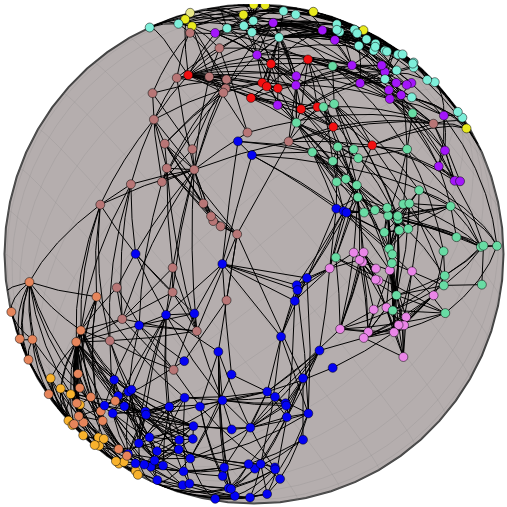}
    \includegraphics[width=0.32\textwidth]{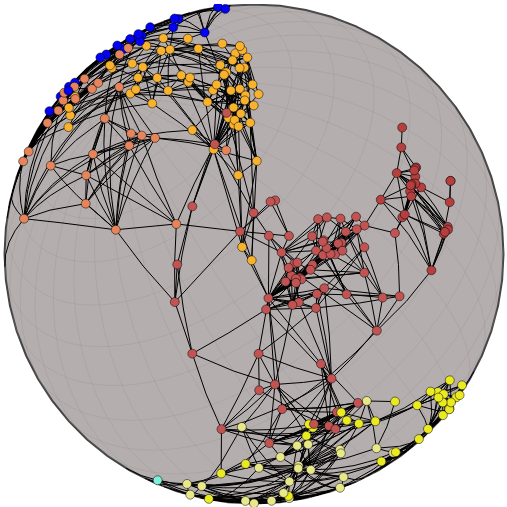}    
    \caption{The Maps of Science dataset~\cite{mapsofscience} laid out using our SMDS algorithm, from three different perspectives. Each color represents a different field of science (nodes are subfields), and their relationships exhibit a ring-like structure. Any field can be placed in the center of the view.}
    \label{fig:map-of-science}
\end{figure}

%Between spherical and hyperbolic geometry, spherical geometry is typically conceptually easier to understand. It can be embedded into 3 dimensional space, and the surface of the Earth is closely approximated by a sphere. 
Spherical geometry has benefits in the context of data visualization. 
In Euclidean (or hyperbolic) layouts, one is forced to choose a ``center" of the embedding, intentionally or not, whereas on the sphere there is no notion of a center.
A perceptual side effect of centered embeddings is that vertices near the center seem more important, while vertices away from the center seem more peripheral. This problem does not occur in spherical space, where simple rotation can place any vertex in the center of the view (a feature that is very useful when visualizing social networks, or networks of research fields); see Fig.~\ref{fig:map-of-science}. 
Additionally, many spherical projections into Euclidean space, such as the stereographic projection, provide a desirable focus+context effect. 
%distorting things near the edge of the projection. 

Some focus+context type algorithms for visualizing large hierarchies by using hyperbolic geometry are discussed in~\cite{lamping1996the,lamping1995}.

Distances in non-Euclidean geometries generalize the concept of a straight line to that of a geodesic, defined as an arc of shortest length (not necessarily unique) that contains both endpoints. The distance between two endpoints is then the length of that curve.

A point on a sphere of radius $R$ is uniquely represented by a pair of angles, $(\phi, \lambda)$, where $0 \leq \phi \leq \pi$ is known as the {\em latitude} and $0 \leq \lambda \leq 2\pi$ is the {\em longitude}. Given two points $(\phi_1, \lambda_1), (\phi_2, \lambda_2)$ on the sphere with radius $R$, the geodesic distance is then derived by the spherical law of cosines:
\begin{equation}
    \label{eq:geodesic}
    \delta((\phi_1,\lambda_1),(\phi_2,\lambda_2)) = R *\arccos( \sin \phi_1 \sin \phi_2 + \cos\phi_1 \cos\phi_2 \cos(\lambda_1 - \lambda_2 ) )
\end{equation}
where $\delta(X_i, X_j)$ denotes the geodesic distance between points $X_i$ and $X_j$, assuming  $X$ is an $n\times 2$ matrix whose rows correspond to spherical coordinates.

\begin{figure}[t]
    \centering
    \includegraphics[width=0.24\textwidth]{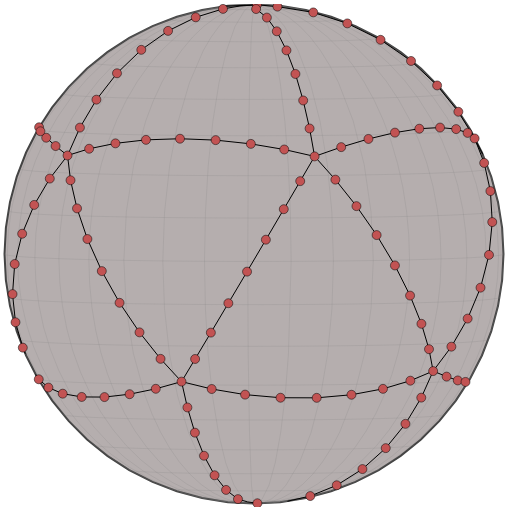}
    \includegraphics[width=0.24\textwidth]{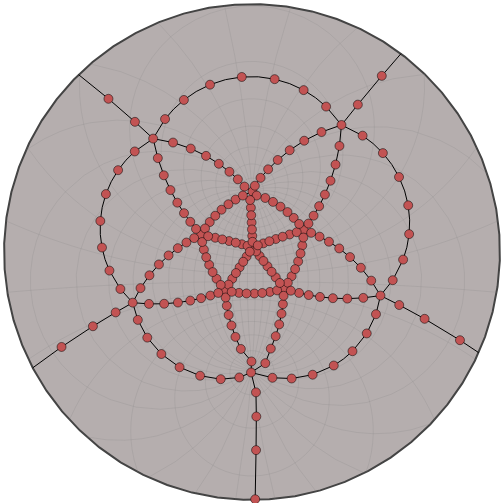}
    \includegraphics[width=0.24\textwidth]{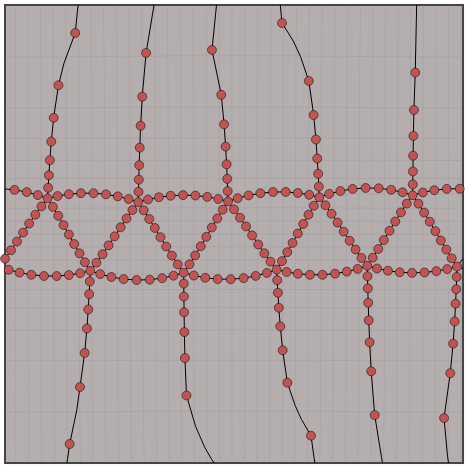}
    \raisebox{0.5\height}{\includegraphics[width=0.24\textwidth]{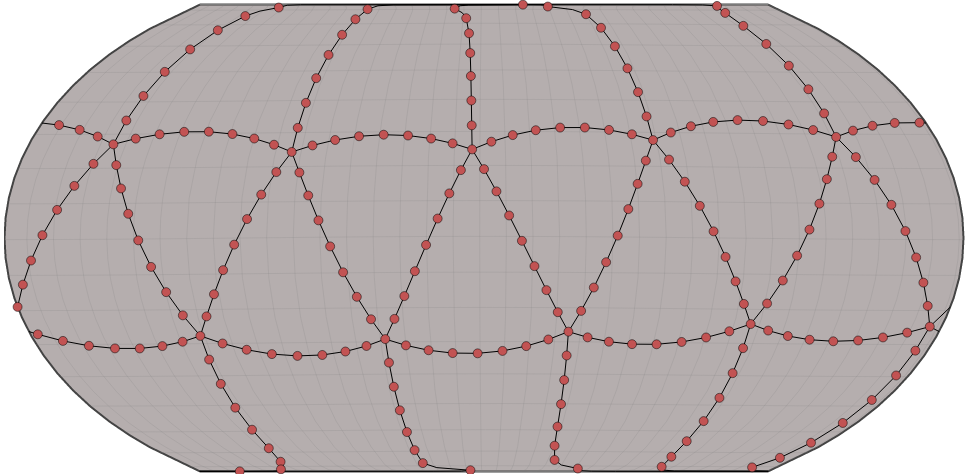}}

    \caption{A subdivided isocahedron graph embedded on the sphere, displayed with the %(from left to right)
    orthographic, stereographic, Mercator, and equal Earth projections.}
    \label{fig:isocahedron}
\end{figure}

It is known that the surface of a sphere cannot be perfectly preserved in any 2-dimensional Euclidean drawing, due to its curvature. One can preserve various combintations of angles, areas, geodesics, or distances but not all of these simultaneously. 
The orthographic projection, or the ``view from space" projects the sphere onto a tangent plane with point of perspective from outside the sphere. While half of the sphere is obscured and shapes and area are distorted near the boundary, geodesics through the origin are preserved and it gives the impression of a 3-dimensional globe. The stereographic projection is similar but instead with a point of projection looking through the sphere, and preserves angles. The Mercator projection is a common cylindrical map projection with heavy area distortion near the poles. The equal Earth projection preserves area and gives the impression of a spherical shape. Examples are shown in Fig.~\ref{fig:isocahedron}. We primarily use the orthographic and equal Earth projections in this paper.

\subsection{Graph Layout in non-Euclidean Geometry}

Non-Euclidean graph visualization has been studied by Munzner, with an emphasis on trees and hierarchies~\cite{munzner1997h3,munzner1998exploring,munzner2000interactive,munzner1995visualizing}, and 
the following link
\url{treevis.net}, provides several examples of hyperbolic and sphere based tree visualizations~\cite{treevis}.
Spherical layouts have been investigated in an immersive setting such as virtual reality~\cite{kwon2016study,DBLP:journals/cgf/YangJDMCC18}. 
Self-organizing maps have been developed for both spherical and hyperbolic geometries~\cite{ontrup2001hyperbolic}. 
Several other examples of spherical graph visualization include the Map of Science~\cite{mapsofscience}, the ``Places and Spaces"~\cite{placesandspaces}, and ``Worldprocessor"~\cite{worldprocessor} exhibitions.
Some limitations of the existing algorithms for hyperbolic graph visualization are discussed in~\cite{eppstein2021limitations}.

%The graph drawing problem is generally defined as determining the position of vertices and routing of edges to effectively draw a node-link diagram. In many cases, the arcs of edges are restricted to straight Euclidean line segments meaning that edge routing is only determined by the position of vertices. Note that these line segments are Euclidean geodesics. When we generalize the graph drawing problem to non-Euclidean geometries, this straight-line restriction of edges instead becomes edges being routed as geodesic segments between vertices. With this restriction, it is sufficient to determine placement of vertices to define a drawing whether we are in Euclidean, hyperbolic, or spherical geometries. 

Force-directed algorithms model the nodes and edges as a physical system, and provide a layout by minimizing the total energy. These algorithms are popular in part due to their conceptual simplicity and quality layouts~\cite{kobourov2012spring}. 
A general technique for generalizing force-directed algorithms to non-Euclidean spaces 
is described in 
~\cite{kobourov2005non}. 
However, it only
works for small graphs as for larger ones it is too computationally expensive and is unlikely to escape local minima.

There are several different approaches to embedding a graph on the sphere. A simple idea is to generate a 2D Euclidean layout and project it onto the sphere through a linear map~\cite{DBLP:conf/chi/DuCLXT17,perry2020drawing}, however, this embedding will not make full use of spherical geometry. 
%Force-directed algorithms can be generalized to Riemannian spaces~\cite{kobourov2005non}, but this is computationally expensive. 
Another approach is to embed the graph in 3D Euclidean space and modify it to force it on the surface of a sphere~\cite{de2009multidimensional,perry2020drawing}, but this is quite mathematically involved and complicates the optimization. 
A more natural method directly computes a 2D spherical embedding (in latitude and longitude) such that the geodesic distances on the sphere and graph-theoretic distances between pairs of vertices are closely matched~\cite{elbaz2005texture}.
We focus on this approach and make it scalable by adopting stochastic gradient descent for the optimization phase and by solving the dilation/resizing problem specified below.
%This allows us to make a simple ``swap" in the objective function from the Euclidean norm between embedded points to the geodesic distance between embedded points, and is the method we use described in section~\ref{sec:algorithm}. 

\begin{figure}[t]
    \centering
    \includegraphics[width=0.32\textwidth]{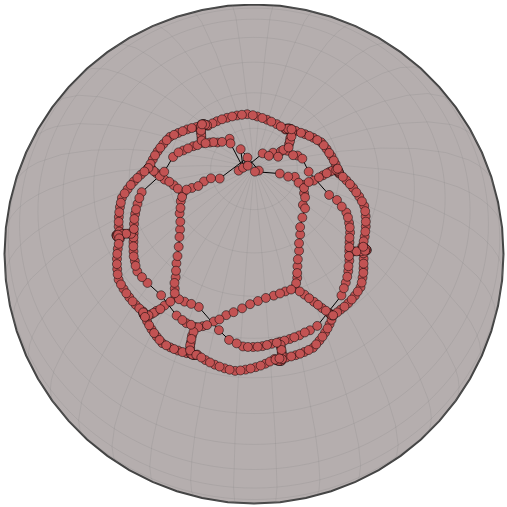}
    \includegraphics[width=0.32\textwidth]{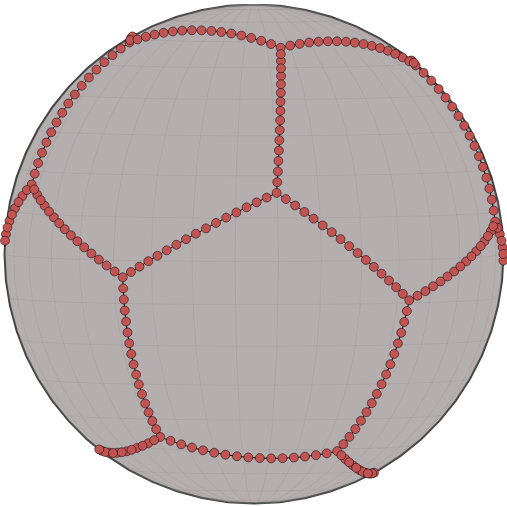}
    \includegraphics[width=0.32\textwidth]{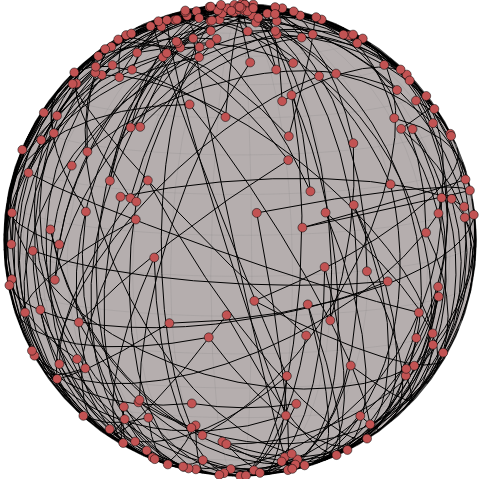}    
    \caption{A dodecahedron subdivision graph. Left: a small dilation factor forces the layout on a small patch of the sphere. Middle: a correct dilation factor using our heuristic discussed in Section~\ref{sec:dilation}, allows the graph to make use of the spherical geometry. Right: a large dilation factor makes the distances unachievable.}
    % unrealizable.}
    \label{fig:dilation-examp}
\end{figure}

In the graph drawing literature, the normalized stress of a layout is a standard quality measure~\cite{DBLP:journals/tvcg/GansnerHN13,DBLP:journals/cgf/KruigerRMKKT17,DBLP:journals/tvcg/ZhuCHHLZ21}.
%, where the layout is resized to minimize the stress function; see~\cite{DBLP:journals/tvcg/GansnerHN13,DBLP:journals/cgf/KruigerRMKKT17,DBLP:journals/tvcg/ZhuCHHLZ21}.
This is perfectly acceptable in Euclidean space where a layout is not meaningfully changed when the layout is resized.
%its size is changed by a positive real number. 
% Unique to non-Euclidean graph layouts is the issue of {\em dilation or resizing}.
For non-Euclidean graph layouts there is a possible issue of {\em dilation or resizing}. 
Formally, a dilation is a function on a metric space $M$,  $f: M \xrightarrow{} M$ that satisfies $d(f(x),f(y)) = rd(x,y)$ for $x,y\in M$, $r > 0 \in \mathbb{R}$ and $d(x,y)$ being the distance between $x$ and $y$.
In non-Euclidean spaces, such as the sphere, the size of a layout can have drastic effects; see Fig.~\ref{fig:dilation-examp}. %Take the surface of the Earth for instance. On small scales it appears flat but over large distances the curvature is noticeable. 
At small dilation, a graph embedded on the sphere takes only a small patch and the sphere patch behaves like a piece of the Euclidean plane. At large dilation, a graph embedded on the sphere wraps over itself. At some optimal dilation the embedded graph fits on the sphere with low distortion.
Choosing the size of the sphere is important to accurately represent the data. We are unaware of any work regarding this problem in spherical embedding, and propose a heuristic and optimization scheme to solve it in Section~\ref{sec:dilation}. 

As stress is difficult to interpret between geometries, we use a more fair comparison metric called \textit{distortion}~\cite{miller2022,sala2018representation} defined later in Section~\ref{sec:geo_compare}.

% The graph embedding problem is that of assigning a set of low dimensional vectors to the vertices of a graph in a way that captures the graph structure. More formally, given a graph $G=(V,E)$ and a dimension $d<<|V|$, find a representation of the graph as a set of $d$-dimensional vectors corresponding to the elements of $V$, which preserves some property or properties as much as possible~\cite{DBLP:journals/tkde/CaiZC18} (e.g., pairwise similarity in MDS). In the general setting, edges can be routed as $d$-dimensional arcs, but here we restrict ourselves to representing edges as straight Euclidean line segments totally determined by the position of their incident edges. When visualizing graphs, the embedding dimension $d$ is usually 2 ( sometimes 3). Thus, to compute an embedding it suffices to find an assignment of 2-dimensional vectors to $V$ that represent the positions of the vertices.

\section{Algorithm}
\label{sec:algorithm}
Our spherical multi-dimensional scaling algorithm (SMDS) resembles that of other stress based graph layout algorithms. 
That is, we first compute a graph-theoretic distance matrix via an all-pairs-shortest-paths algorithm and then use this distance matrix 
as an input
to minimize the generalized stress function and compute vertex coordinates on the sphere.
This differs from standard Euclidean MDS in that Euclidean distances between points are replaced by geodesic distances
between the points on sphere. 
The corresponding stress function defined on the sphere is
\begin{equation}
    \label{eq:gen_stress}
    \sigma_S(X) = \sum_{i<j} w_{ij} ( \delta(X_i, X_j) - d_{ij} )^2
\end{equation}
Where $\delta(X_i, X_j)$ denotes the geodesic distance between vertices $i$ and $j$, $d_{ij}$ is the graph-theoretic distance between vertex $i$ and $j$, and $w_{ij}$ is a weight, typically set to $d_{ij}^{-2}$. 
However, one can also give preferred weights based on the importance of the points and based on the application. 
Another typical choice is binary weights, where $w_{ij}$ is either $0$ or $1$.
Unless otherwise specified, $\delta$ corresponds to the geodesic distances on the unit sphere and $\delta_R$ the geodesic distances on a sphere with radius $R$.

\begin{figure}[t]
    \centering
    \includegraphics[width=0.4\textwidth]{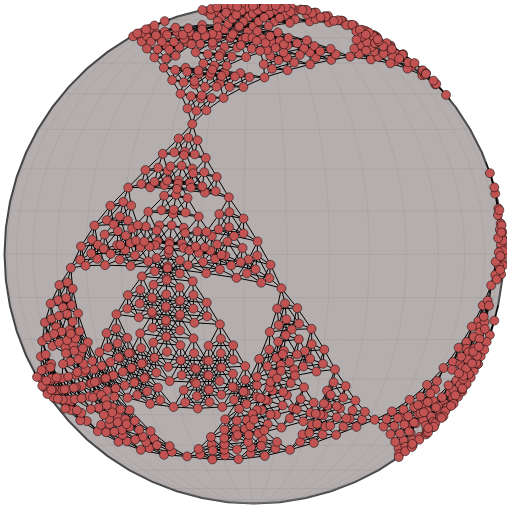}\hspace{.5cm}
    \includegraphics[width=0.4\textwidth]{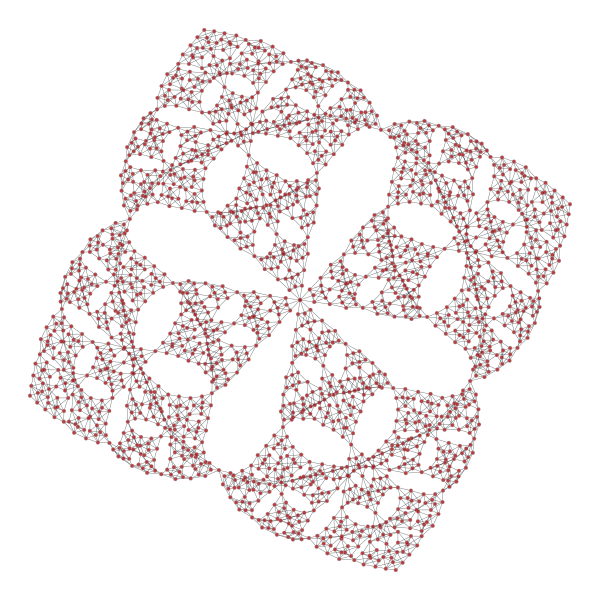}

    \caption{The Sierpinski3d~\cite{DBLP:journals/cgf/KruigerRMKKT17} graph on the sphere (left) and in the plane (right). While the Euclidean drawing on the right is aesthetically pleasing, it looks deceptively like a 2D structure and implies a center. The sphere more accurately captures the structure.
    %, showing that there is no center after all.
    }
    \label{fig:sierpinski}
\end{figure}
We minimize equation~\eqref{eq:gen_stress} by stochastic gradient descent (SGD), which we experimentally show converges in fewer iterations while achieving lower distortion than exact gradient descent for sufficiently large graphs.
%than other methods for graph drawing such as gradient descent or stress majorization in Euclidean geometry~\cite{DBLP:conf/gd/BorsigBP20,DBLP:journals/tvcg/ZhengPG19}, and for gradient descent in hyperbolic geometry~\cite{miller2022}.
SGD is a minimization approach in the gradient descent family of algorithms. Fully computing the exact gradient can be too expensive %($O(|V|^2)$ for stress). 
and SGD instead repeatedly performs a constant time approximation of the gradient,
by considering only a single term of the sum (or subset of terms for mini-batch stochastic),
which allows it to make more updates. 
%than the exact gradient descent.
As a result, SGD tends to converge in fewer iterations while more consistently finding the global minimum~\cite{ruder16}.

To perform SGD on the stress function, we approximate the gradient by looking at only a single pair of vertices. Note that this corresponds to one summand of the full stress function. If we rewrite equation~\eqref{eq:gen_stress} as $\sigma_S(X) = \sum_{i<j}f_{i,j}(X)$ then we can more simply write the full gradient in terms of $f$.
Apply the chain rule to see we will need to derive the partial gradient of the geodesic distance function:

\[\frac{\partial \delta (X_i, X_j)}{\partial \phi_i} = \frac{-\cos \phi_i \sin\phi_j - \sin\phi_i \cos\phi_j \cos(\lambda_i - \lambda_j)}{\sqrt{1 - \cos^2 \left( \delta (X_i, X_j)\right)}}\]

\[\frac{\partial \delta (X_i, X_j)}{\partial \lambda_i} = \frac{\cos\phi_i \cos\phi_j \sin(\lambda_i - \lambda_j)}{\sqrt{1 - \cos^2 \left(\delta (X_i, X_j)\right)}}\]

Unlike in Euclidean space, the gradient of the spherical distance function %in general
is not symmetric, i.e.,  $\frac{d\delta(X_i, X_j)}{d(\phi_i ,\lambda_i)} \neq - \frac{d\delta (X_i, X_j)}{d(\phi_j ,\lambda_j)}$. Writing out the full gradient:

\begin{equation}
\label{eq:grad}    
\frac{\partial f_{i,j}}{\partial X_k} = 
    \begin{cases}
    2w_{i,j}(\delta (X_i, X_j) - d_{ij}) \left( \frac{\partial \delta (X_i, X_j)}{\partial \phi_i},\frac{\partial \delta (X_i, X_j)}{\partial \lambda_i} \right) &  k = i \\
    2w_{i,j}(\delta (X_i, X_j) - d_{ij}) \left( \frac{\partial \delta (X_i, X_j)}{\partial \phi_j},\frac{\partial \delta (X_i, X_j)}{\partial \lambda_j} \right) &  k= j \\
    0 & \text{otherwise}
    \end{cases}
\end{equation}

We can apply SGD to equation~\eqref{eq:gen_stress} by selecting pairs $i,j$ in random order, and updating $X$ by $X - \eta \frac{\partial f_{i,j}}{\partial X_k}$ where $\eta$ is the learning rate; see Algorithm~\ref{alg:sgd}.

We randomly initialize the placement of vertices uniformly on the sphere, as other work has shown that SGD is consistent across initialization strategies~\cite{DBLP:conf/gd/BorsigBP20,miller2022,DBLP:journals/tvcg/ZhengPG19}.

% \begin{figure}
%     \centering
%     \includegraphics[width=0.4\textwidth]{figures/polytopes/cube_4_ortho.png}
%     \includegraphics[width=0.4\textwidth]{figures/polytopes/Dodecahedron_4_ortho.png}
    
%     \includegraphics[width=0.4\textwidth]{figures/polytopes/cube_4_euclid.png}
%     \includegraphics[width=0.4\textwidth]{figures/polytopes/dodecahedron_4_euclid.png}    
%     \caption{A subdivided cube (left) and dodecahedron (right) laid out by SMDS (top) and Euclidean MDS (bottom). Not only do the spherical layouts have lower distortion (see table **insert) but they avoid implying a central vertex.}
%     \label{fig:polytopes}
% \end{figure}

\begin{algorithm}
\caption{Stochastic gradient descent algorithm for spherical MDS}
\label{alg:sgd}
\begin{algorithmic}
\Procedure{Stochastic gradient descent}{d}
\State $X \gets $random initialization
\While{ $\Delta(\text{stress}(X)) > \epsilon$ or max iterations is reached}
\For{(i,j) in random order}
    \State $X_i \gets X_i-\eta \frac{d f_{i,j}}{d X_i}$, according to \eqref{eq:grad} 
    \State $X_j \gets X_j-\eta \frac{d f_{j,i}}{d X_j}$, according to \eqref{eq:grad}
\EndFor
\EndWhile
\State \textbf{return} $X$
\EndProcedure
\end{algorithmic}
\end{algorithm}

\section{Evaluation}
We first investigate the various parameters that effect SGD's optimization, then compare our results to exact gradient descent. 

\subsection{Hyper-parameters}
\label{sec:param}
There are several hyper-parameter choices to be made when using SGD. The learning rate $\eta$ (also known as step size, annealing rate) has a large effect on the resulting embedding. If the learning rate is too large, the optimization will ``overstep" and either fluctuate around a minimum, or diverge.
%either way it will not converge to a minimum. 
If the learning rate is too small, the optimization may require many iterations to converge and is more likely to converge to a local minimum. A better strategy is to employ a learning rate schedule, where at early iterations the learning rate is large but decreases over time to allow for convergence. This is known to converge to a stationary point (could be a saddle) under certain conditions: $\sum g(t) = \infty$ and $\sum g(t)^2 < \infty$~\cite{bottou2012stochastic}.

\begin{figure}[t]
    \centering
    \includegraphics[width=0.49\textwidth]{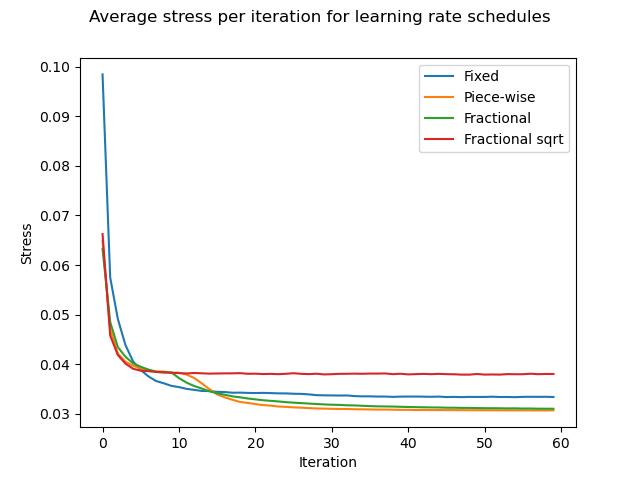}
    \includegraphics[width=0.49\textwidth]{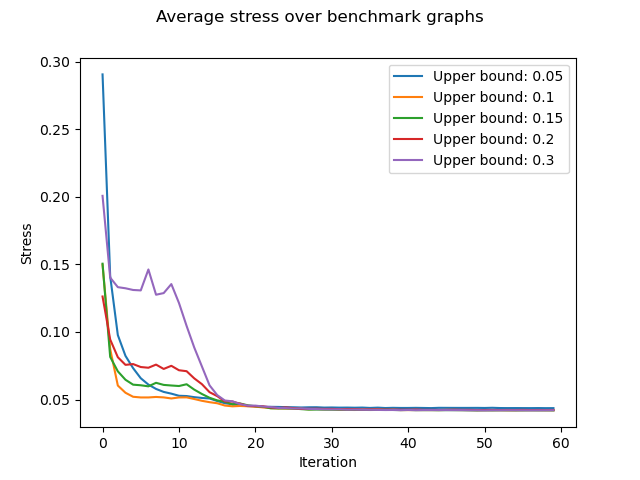}
    \caption{(Left) Effect of the learning rate schedule on the optimization.
    The piece-wise schedule adapted from~\cite{DBLP:journals/tvcg/ZhengPG19} arrives at a minimum faster on average.
    (Right) Effect of the upper bound on the learning rate on the optimization. An upper bound of 0.1 behaves predictably.
    Values for both are averaged over all graphs in our benchmark.}
    \label{fig:upperbound}
\end{figure}

We investigate a limited subset of possible learning rate schedules, a fixed learning rate at 0.05, a piece-wise schedule similar to that of Zheng et al.~\cite{DBLP:journals/tvcg/ZhengPG19}, a fractional decay of $\Theta(t^{-1})$, and a slower fractional decay of $\Theta(t^{-.5})$. The piece-wise schedule begins with an exponential decay function, with large initial values and switches to $\Theta(t^{-1})$ once below a threshold. There are a few changes we needed to make to the piece-wise schedule. Firstly, while Zheng et al.~\cite{DBLP:journals/tvcg/ZhengPG19} upper bound their learning rate by 1, this upper bound is too large for SMDS. The upper bound for the Euclidean algorithm was derived from the geometric structure, and a value of 1 reduces the stress between a single pair of vertices to 0. The latitude and longitude of the sphere are angles and so do not have this property. We instead need a relatively small upper bound, noting that large movements of a pair vertices on the sphere that need to be moved apart can actually bring them closer together (by wrapping around the sphere). We investigated values in the range 0 to 1, and settled on 0.1 as it achieves low stress quickly while not being so small as to fall into local minima; see Fig.~\ref{fig:upperbound}.

% \begin{figure}
%     \centering
%     \includegraphics[width=0.40\textwidth]{figures/learning_rate/494_bus.png}
%     \includegraphics[width=0.40\textwidth]{figures/learning_rate/cube_4.png}
    
%     \includegraphics[width=0.40\textwidth]{figures/learning_rate/dwt_918.png}
%     \includegraphics[width=0.40\textwidth]{figures/learning_rate/dwt_419.png}
%     \caption{The effect of different learning rate schedules on the stress for a selection of graphs. The piece-wise learning rate described in section~\ref{sec:param} consistently achieves a low minimum more quickly than other schedules.}
%     \label{fig:my_label}
% \end{figure}

Randomization is a to select pairs $i,j$ in the stochastic optimization function. While SGD was originally done using sampling with replacement, random reshuffling has been shown to converge in fewer total updates~\cite{gurbuzbalaban2021random}. To use random reshuffling in stress minimization, we enumerate all pairs $i<j$ of vertices in a ordered list and shuffle this list after every iteration. This ensures that the order in which we visit pairs is random, but that each pair is visited before we sample the same pair again.

A stopping condition is how the algorithm determines to terminate, either by converging or by reaching a maximum number of iterations. We measure convergence by tracking the change in the value of the optimization function between iterations. In the figures and evaluation results below, we set the convergence threshold to $1e^{-7}$ or a balance between speed and quality.

\subsection{Evaluation}
Our code is open source and written in Python. Experiments are performed on an Intel® Core™ i7-3770 CPU @ 3.40GHz × 8 with 32 GB of RAM with a 64 bit installation of Ubuntu 20.04.3 LTS.

We use a set of 40 graphs to evaluate our SMDS algorithm: 34 from the 
\textit{SuiteSparse Matrix Collection}~\cite{DBLP:journals/toms/DavisH11} and the remaining 6 from skeletons of 3-dimensional polytopes. We use the cube, dodecahedron, and isocahedron, and subdivide them 4 times each to obtain cube\_4, dodecahedron\_4 and isocahedron\_4. 
We present spherical layouts of a subset of our benchmark graphs; see Table~\ref{fig:layouts}. The remaining layouts can be found in the Appendix. We see that there are several graphs particularly well suited to spherical layout: the 3D  polytopes and their subdivisions have much lower distortion on the sphere than in the plane. 3-dimensional meshes and triangulations of surfaces such as dwt\_307 and delaunay\_n10 also have lower error on the sphere.

\begin{figure}
    \centering
    \includegraphics[width=0.49\textwidth]{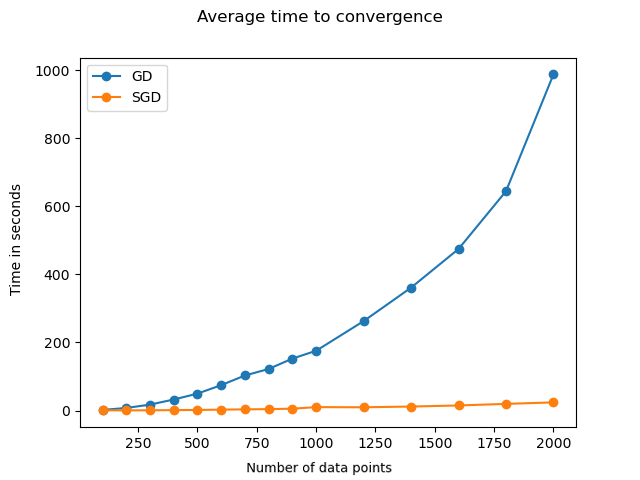}
    \includegraphics[width=0.49\textwidth]{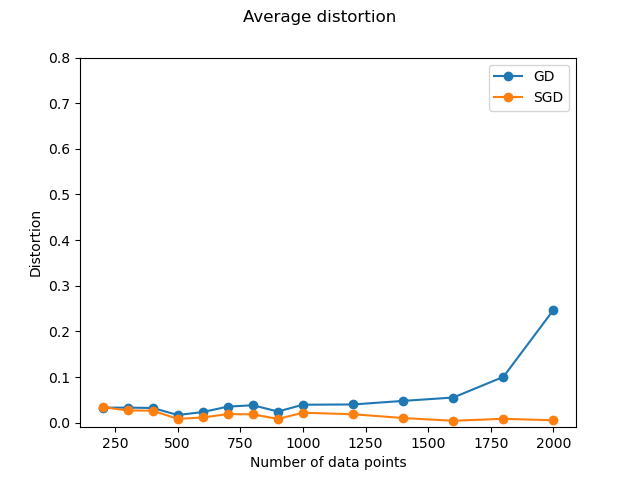}
    \caption{How the SGD optimization scheme fairs compares to the exact GD in terms of time (left) and error (right). The larger the size of the graph, the more benefit is seen from the use of SGD.}
    \label{fig:time_error}
\end{figure}

The SGD optimization scheme performs better than exact GD 
on both time to convergence and stress as the size of the data becomes large as we expect; see Fig.~\ref{fig:time_error}.

\begin{table*}[htp]
  \caption{Layouts} 
  \label{fig:layouts}
  \centering
  \begin{tabular}
      {l c c |l  c c} \hline & E-MDS & SMDS &  & E-MDS & SMDS  \\
      \hline 
      
      \raisebox{-.5\normalbaselineskip}[0pt][0pt]{\rotatebox[origin=c]{90}{can\_73}} &
      \parbox[c]{\tablesize\textwidth}{
      \includegraphics[width=\tablesize\textwidth]{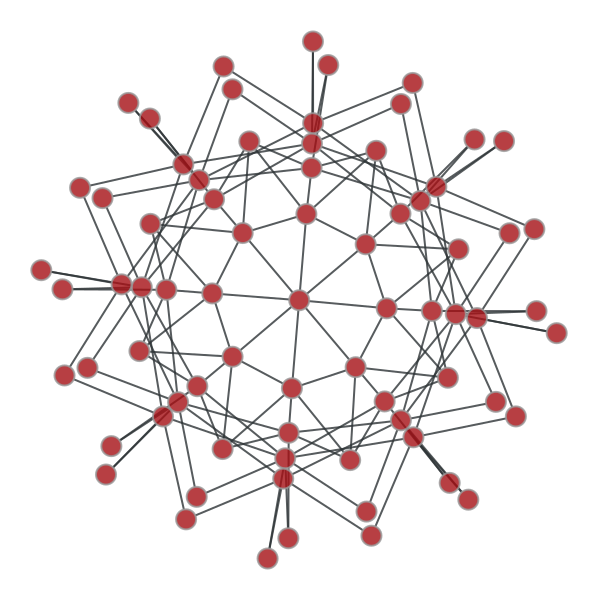}} 
      & \parbox[c]{\tablesize\textwidth}{
      \includegraphics[width=\tablesize\textwidth]{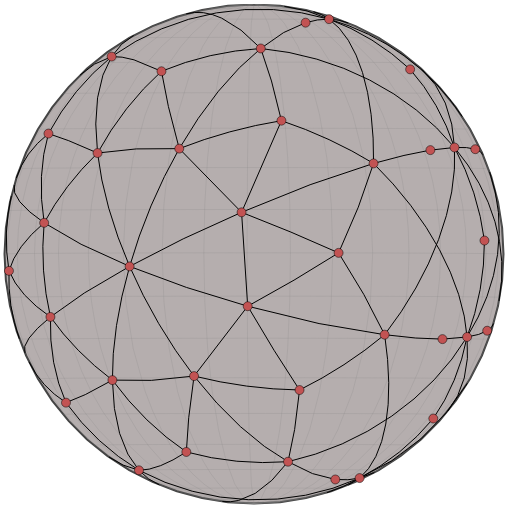}}
      &
      \raisebox{-.5\normalbaselineskip}[0pt][0pt]{\rotatebox[origin=c]{90}{can\_96}} &
      \parbox[c]{\tablesize\textwidth}{
      \includegraphics[width=\tablesize\textwidth]{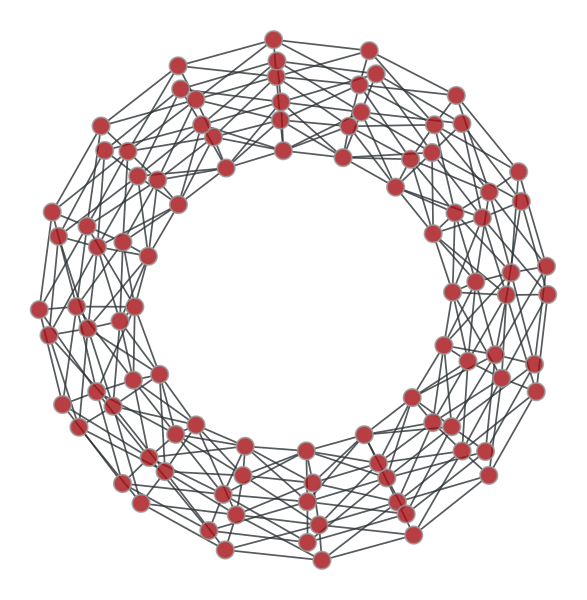}} 
      & \parbox[c]{\tablesize\textwidth}{
      \includegraphics[width=\tablesize\textwidth]{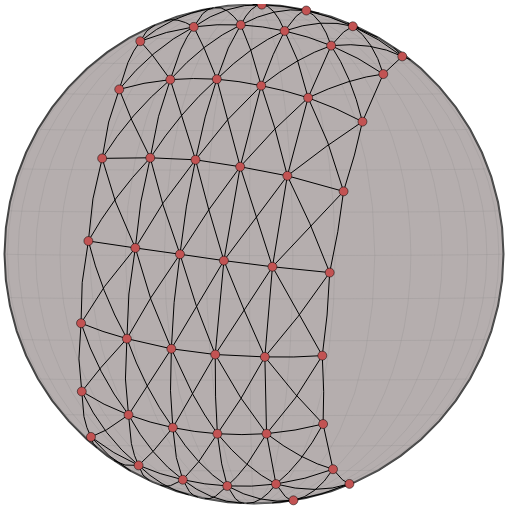}} \\
      
      \raisebox{-.5\normalbaselineskip}[0pt][0pt]{\rotatebox[origin=c]{90}{rajat\_11}} &
      \parbox[c]{\tablesize\textwidth}{
      \includegraphics[width=\tablesize\textwidth]{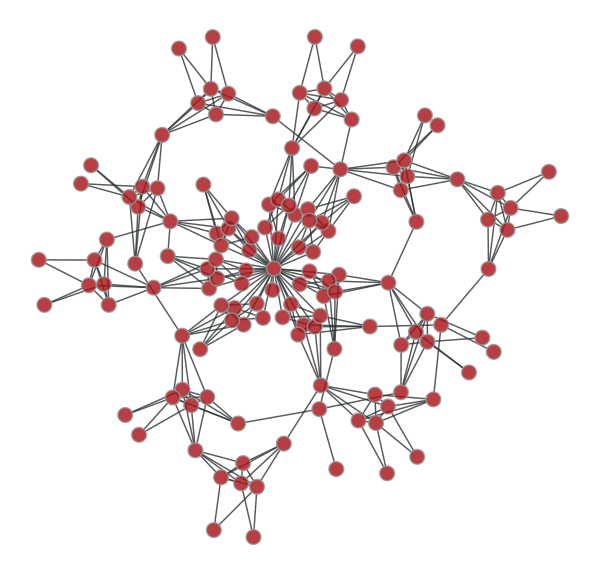}} 
      & \parbox[c]{\tablesize\textwidth}{
      \includegraphics[width=\tablesize\textwidth]{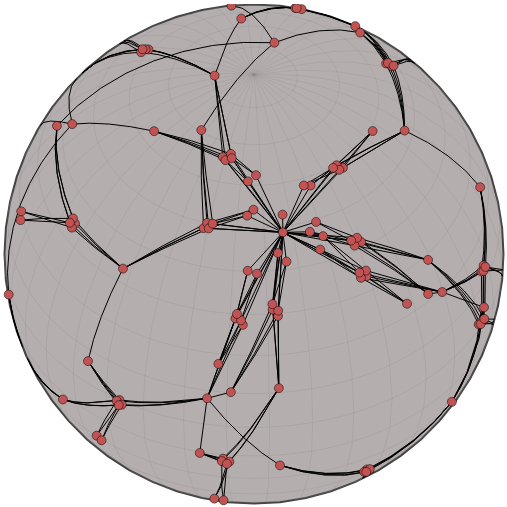}}
    &
      \raisebox{-.5\normalbaselineskip}[0pt][0pt]{\rotatebox[origin=c]{90}{bcsstk09}} &
      \parbox[c]{\tablesize\textwidth}{
      \includegraphics[width=\tablesize\textwidth]{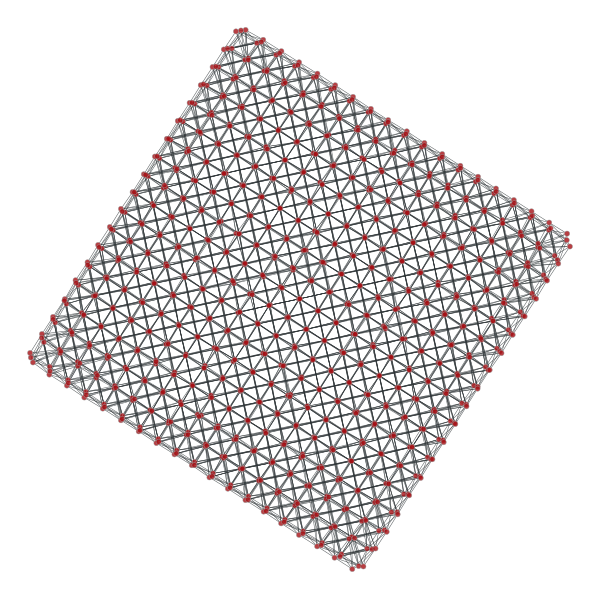}} 
      & \parbox[c]{\tablesize\textwidth}{
      \includegraphics[width=\tablesize\textwidth]{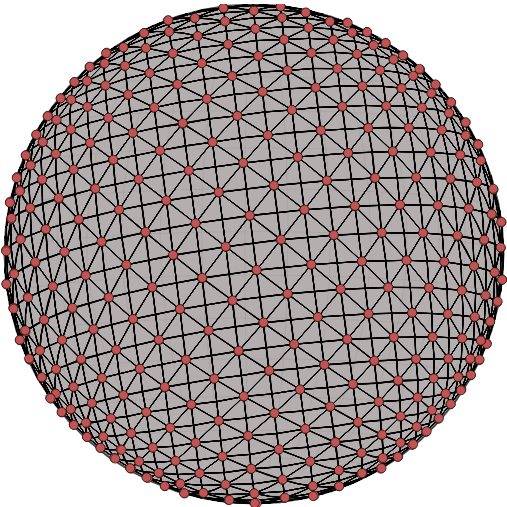}}\\
      
      \raisebox{-.5\normalbaselineskip}[0pt][0pt]{\rotatebox[origin=c]{90}{can\_161}} &
      \parbox[c]{\tablesize\textwidth}{
      \includegraphics[width=\tablesize\textwidth]{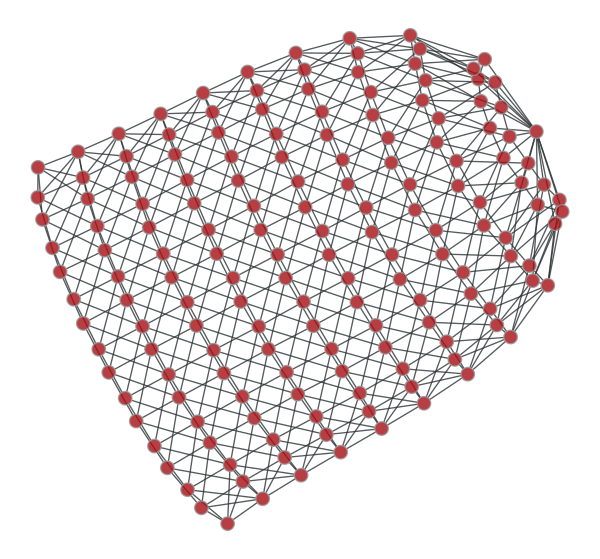}} 
      & \parbox[c]{\tablesize\textwidth}{
      \includegraphics[width=\tablesize\textwidth]{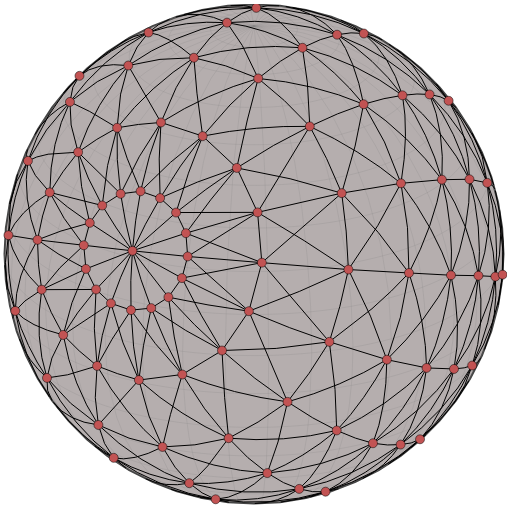}}
    &
        \raisebox{-.5\normalbaselineskip}[0pt][0pt]{\rotatebox[origin=c]{90}{cube\_4}} &
      \parbox[c]{\tablesize\textwidth}{
      \includegraphics[width=\tablesize\textwidth]{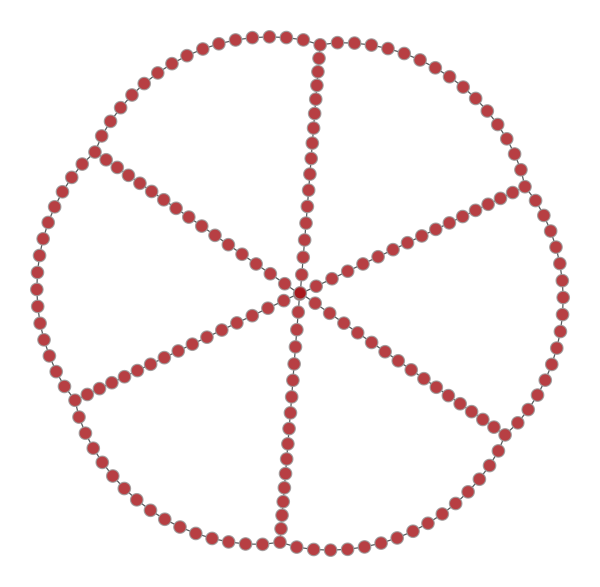}} 
      & \parbox[c]{\tablesize\textwidth}{
      \includegraphics[width=\tablesize\textwidth]{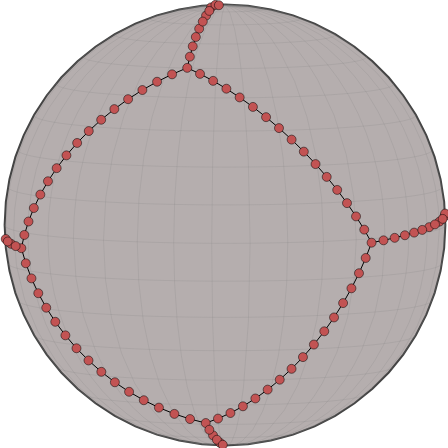}}\\
      
      \raisebox{-.5\normalbaselineskip}[0pt][0pt]{\rotatebox[origin=c]{90}{grid17}} &
      \parbox[c]{\tablesize\textwidth}{
      \includegraphics[width=\tablesize\textwidth]{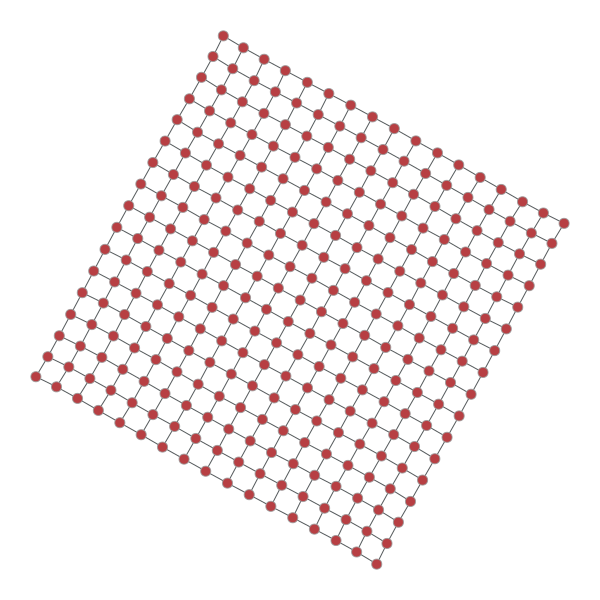}} 
      & \parbox[c]{\tablesize\textwidth}{
      \includegraphics[width=\tablesize\textwidth]{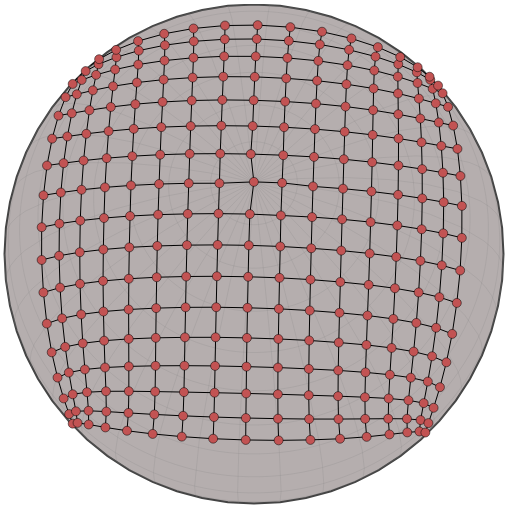}}
    &
      \raisebox{-.5\normalbaselineskip}[0pt][0pt]{\rotatebox[origin=c]{90}{block\_400}} &
      \parbox[c]{\tablesize\textwidth}{
      \includegraphics[width=\tablesize\textwidth]{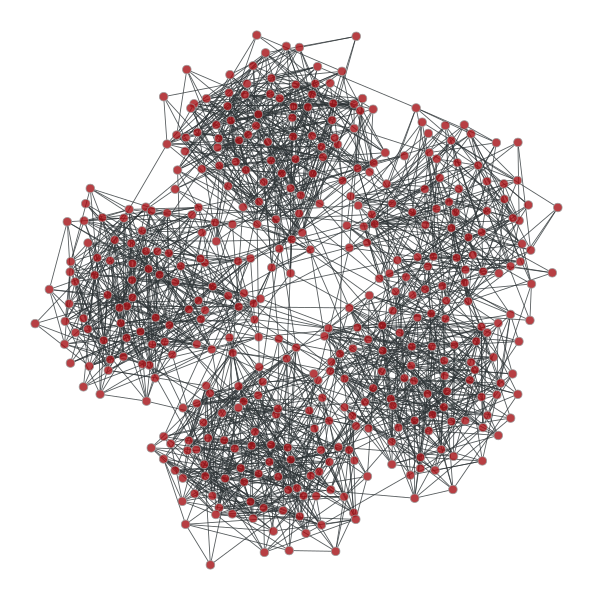}} 
      & \parbox[c]{\tablesize\textwidth}{
      \includegraphics[width=\tablesize\textwidth]{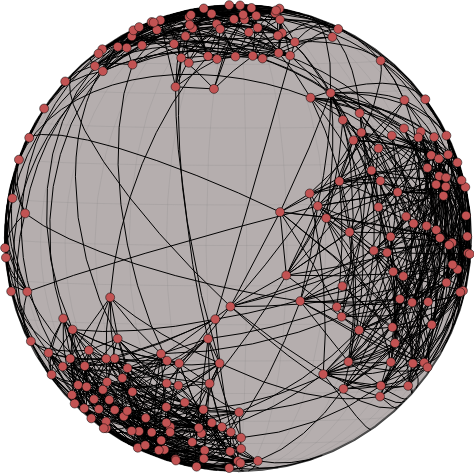}} \\       

      \hline
  \end{tabular}
  
\end{table*}

% \subsection{General Data}
% Our proposed algorithm is not limited to only graph data.
% In this section we use the spherical MDS to visualize MNIST handwritten digit database~\cite{lecun-mnisthandwrittendigit-2010} on sphere.
% The MNIST is a dataset that contain $70,000$ handwritten digits in greyscale, each of sizes $28 \times 28$.

% To apply SMDS algorithm to the MNIST dataset, we first vectorize the matrices that correspond to the greyscale pixel values of each image. Thus, each datapoint, corresponding to one greyscale image of a digit, is a vector of size $28 \times 28 = 784$. 
% Thus, the dataset that we start with contains $70000$ datapoints in $\mathbb{R}^{784}$. 
% We sample 1000 points from this set, and use these datapoints to compute the distances between each pair and feed the matrix of distances to the spherical MDS algorithm. The achieved embedding is presented in Fig.~\ref{fig:mnist}. We used the same colors for the same digits.

% \begin{figure}
%     \centering
%     \includegraphics[width=\textwidth]{figures/mnist.png}
%     \caption{The embedding of MNIST dataset on the sphere by SMDS algorithm. Same colors correspond to the images of the same digits.}
%     \label{fig:mnist}
% \end{figure}

\section{Geometry Comparison}
\label{sec:geo_compare}
Here we discuss some possible drawbacks of graph embedding in spherical space and compare graph embeddings between Euclidean, spherical and hyperbolic spaces.
Stress works well for producing layouts, but directly comparing stress scores between geometries are difficult to interpret. Layouts are often uniformly scaled so that stress is minimum before reporting (see~\cite{DBLP:journals/tvcg/GansnerHN13,DBLP:journals/cgf/KruigerRMKKT17}) which works fine in Euclidean space, but becomes a problem in spherical and hyperbolic spaces.
In order to more fairly compare embedding error across geometries, we use the \textit{distortion}~\cite{miller2022,sala2018representation} metric, defined as 
\begin{equation}
    \label{eq:distortion}
    \text{distortion} (X) = \frac{1}{{|V|\choose 2}}\sum_{i<j} \frac{|\delta(X_i, X_j) - d_{i,j}|}{d_{i,j}}.
\end{equation}

\subsection{Dilation of Distances }
\label{sec:dilation}

\begin{figure}
    \centering
    \includegraphics[width=0.3\textwidth]{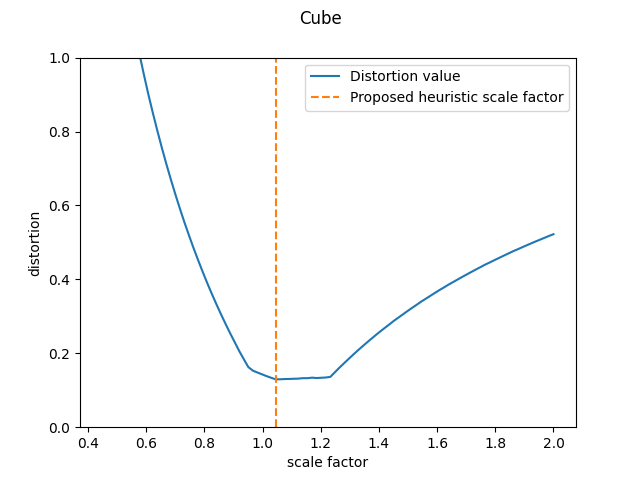}
    \includegraphics[width=0.3\textwidth]{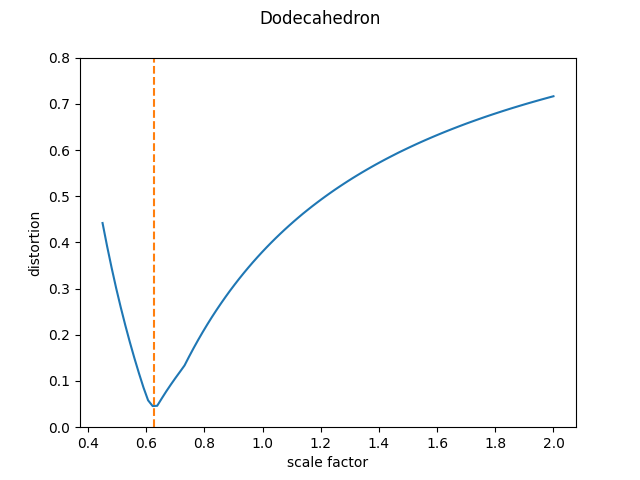}
    \includegraphics[width=0.3\textwidth]{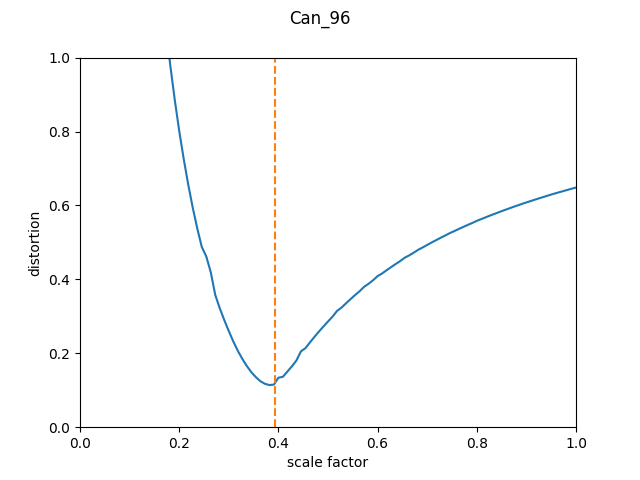}    
    \caption{Effect of dilation on distortion. Our proposed heuristic (orange line) is often very close to the minimum (of the blue curve).
    %In all data we studied, there appears to be a unique minimum.
    }
    \label{fig:opt-v-heur}
\end{figure}

It is known that Euclidean MDS is invariant to dilation, that is
if one multiplies the given distances by a positive real number, the corresponding MDS solution is the original MDS solution multiplied by the same scalar factor (up to rotation). However, this is not true for spherical and hyperbolic spaces. Moreover, 
% A major difference of embedding a graph in Euclidean space and on sphere is that 
 spherical space is bounded, unlike Euclidean space. 
For example, on the 2D unit sphere the maximum distance 
that can be achieved
between two points is $\pi$ (assuming that between any two points we always take the shortest geodesic distance). 
Any graph with diameter (longest shortest path) longer than $\pi$ cannot possibly be embedded on the unit sphere with zero error. 
%Given a graph with graph theoretic distances between the vertices, a reasonable embedding on the sphere should have the possibility to place points on the sphere such that any given distance is achievable.
A reasonable solution is to dilate the input distances so that all the given distances are less than or equal to $\pi$. That is, to multiply the distance matrix, $d$, by $\frac{\max d}{\pi}$. This heuristic appears to work well in practice; see Fig.~\ref{fig:opt-v-heur}. For all of our experiments and layouts, we use this heuristic.
However, this has no guarantees of being optimal.

\begin{figure}
    \centering
    \includegraphics[width=0.32\textwidth]{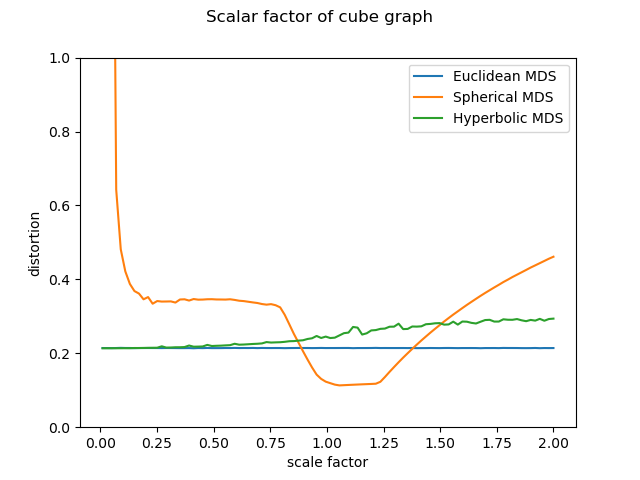}
    \includegraphics[width=0.32\textwidth]{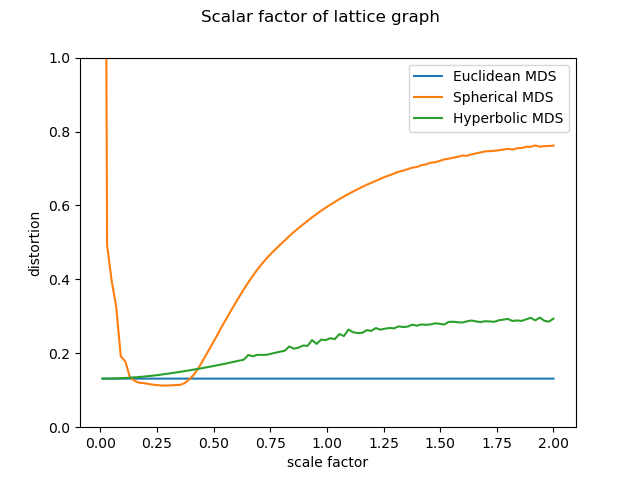}
    \includegraphics[width=0.32\textwidth]{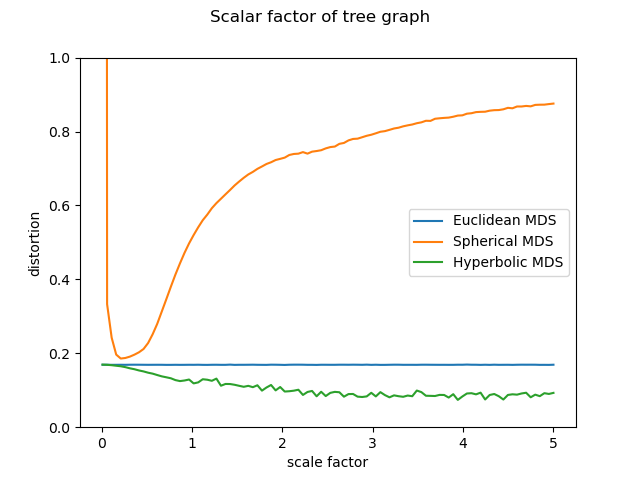}

    \caption{Behavior of distortion on selected graphs with respect to dilation factor in each geometry.}
    \label{fig:dilation_experiment}
\end{figure}

One possible approach to the dilation problem is to make the radius of the sphere also a parameter to optimize. The problem would then become finding the best radius so that the defined stress function is as small as possible.
This can be captured by reformulating Eq.~\eqref{eq:gen_stress} to also optimize the radius:
\begin{equation}
    \label{eq:sphereMDS_scale}
    \arg \min_{R, X_1, \dots X_n \in S_{R}^2} \sum_{i, j = 1}^N \left( \delta_{R} (X_i, X_j) - d_{ij} \right)^2.
\end{equation}
Here $\delta_{R} (X_i, X_j)$ corresponds to the geodesic distance on the sphere with radius $R$ between points $X_i$ and $X_j$. We derive the gradient for $R$ and update it along with the vertex positions at each update step.
%We compare this optimization scheme to our proposed heuristic; see Fig.~\ref{fig:opt-v-heur}.

% % We remark that scaling the distances by the same constant is equivalent to keeping the distances and changing the radius of the sphere.
% % An equivalent formulation to \eqref{eq:sphereMDS_scale} is
% % \begin{equation}
% %     \label{eq:sphereMDS_scale_1}
% %     \arg \min_{R, X_1, \dots X_n \in S_{R}^2} \sum_{i, j = 1}^N \left( \delta (X_i, X_j) - \frac{d_{ij}}{R} \right)^2.
% % \end{equation}
% where we consider the unit sphere and instead dilate the dataset by a factor $R$. 
To the best of our knowledge none of the existing algorithms for spherical embedding consider this dilation/resizing problem. However, we believe that it is a crucial parameter while embedding/drawing a graph on the sphere. 
%We experimentally show that depending on this value the distortion value significantly changes. In Fig.~\ref{fig:dilation_experiment} we demonstrate how dilating the distances changes the distortion value.

\subsection{Choosing Between Geometries}
\label{sec:geo_choose}

% \begin{figure}
%     \centering
%     \includegraphics[width=0.4\textwidth]{figures/short_dist_tab.png}
%     \caption{A subset of results from our direct comparison of geometry experiments.}
%     \label{fig:shot-distortion-tab}
% \end{figure}

\begin{figure}
    \centering
    \includegraphics[width=0.42\textwidth]{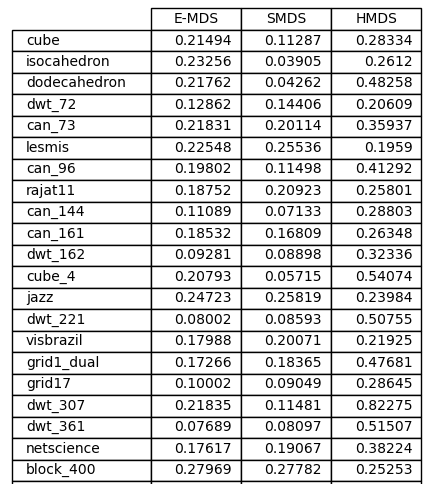}
    \includegraphics[width=0.54\textwidth]{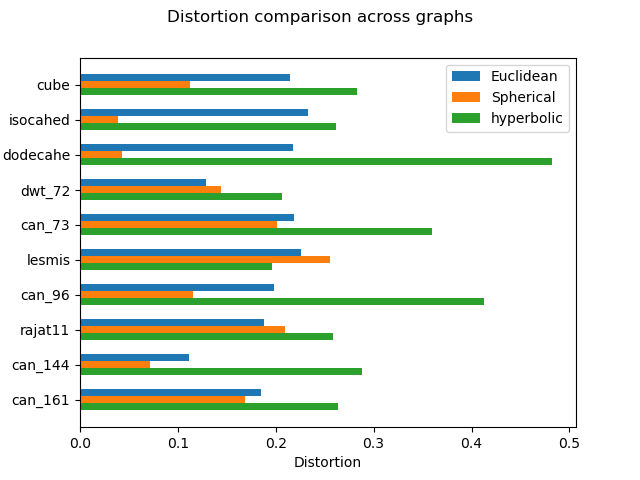}
    \caption{The left subfigure shows a subset of results from the direct comparison for distortion in Euclidean, spherical and hyperbolic space. The right subfigure plots the first 10 rows. We note that 3D polytopes and meshes (the can graphs) are particularly well suited to the sphere, the LesMis graph is a complex network which is best embedded into hyperbolic space, and Euclidean space is better for the remaining ones.}
    \label{fig:shot-distortion-tab}
\end{figure}

One reason to consider embedding graphs on different manifolds (Euclidean, hyperbolic, spherical) is to be able to preserve and visualize important properties of the given graph. Some graphs achieve lower distortion on the sphere, others in hyperbolic space. %however, it is not obvious which manifold to select for an unknown graph. 
In this section we 
investigate how spherical graph layouts differ from other consistent geometries. 
We choose a selection of graphs from the sparse matrix collection, and lay them each out using the Euclidean, spherical, and hyperbolic variants of MDS and measure the distortion. 
We repeat the layout 5 times each, and report the average distortion for each graph in each geometry. We make use of~\cite{DBLP:journals/tvcg/ZhengPG19} for the Euclidean MDS implementation and~\cite{miller2022} for the hyperbolic MDS (HMDS) implementation. 
%Our results are shown Fig.~\ref{fig:shot-distortion-tab}, with additional data in  Appendix~\ref{sec:appendix-geo-compare}. 

% \begin{figure}
%     \centering
%     \includegraphics[width=0.75\textwidth]{figures/across_geo/Dist_across_geo2.png}
    
%     %\includegraphics[width=0.75\textwidth]{figures/Dist_across_geo1.png}
%     \caption{Direct comparison of distortion values between geometries on different graphs. 3-dimensional polytopes and meshes (the can graphs) are particularly well suited to the sphere and so have low distortion. The lesmis graph is a complex network which can be embedded into hyperbolic space with lower distortion. The remaining graphs are best drawn in Euclidean space.}
%     \label{fig:across_geo}
% \end{figure}

The hypothesis we test here is that
some graphs have a dramatically lower distortion in a particular geometry. 
For instance, rectangular lattices can be embedded with constant error in Euclidean space~\cite{verbeek2016}, regular 3D polytopes can be thought of as tesselations of the sphere, and trees have been described as ``discrete hyperbolic spaces"~\cite{krioukov2010hyperbolic}.
The results are summarized in Fig.~\ref{fig:shot-distortion-tab} with additional data in  Appendix~\ref{sec:appendix-geo-compare}. 
We observe that spherical geometry is in fact able to embed polytopes and 3D meshes with lower distortion. Further, hyperbolic geometry is able to embed networks with ``small-world" properties such as lesmis and block\_400 with lower distortion. In graphs with 2D structure, Euclidean space is the clear winner. 
\begin{figure}
    \centering
    \includegraphics[width=0.3\textwidth]{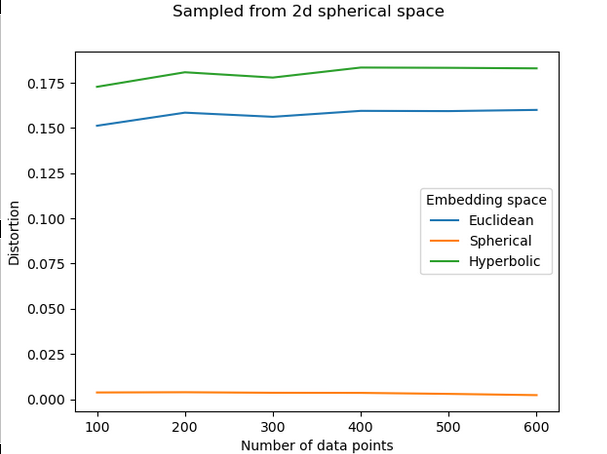}
    \includegraphics[width=0.3\textwidth]{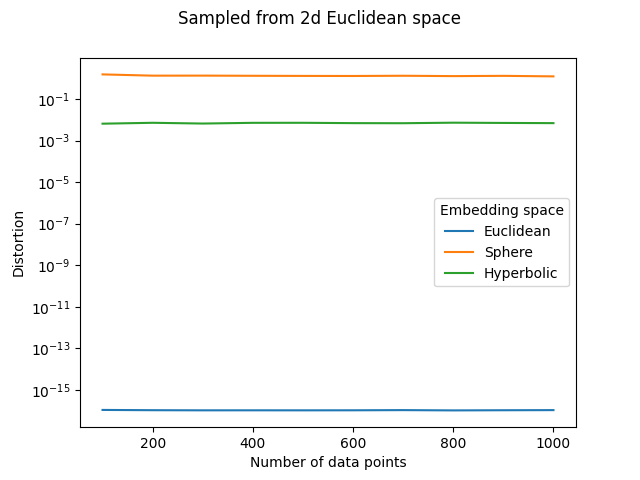}
    \includegraphics[width=0.3\textwidth]{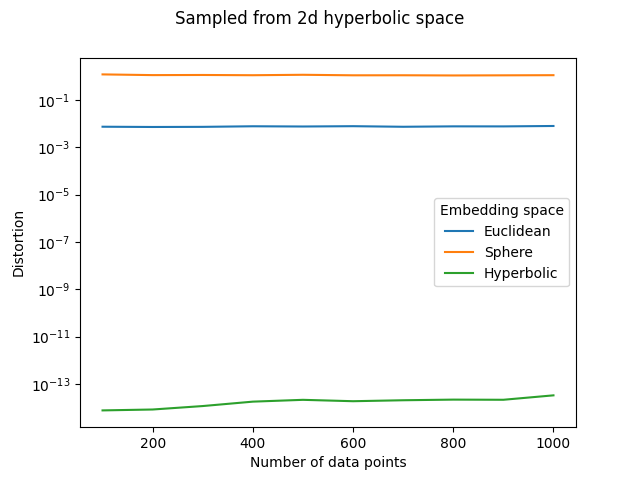}    
    \caption{Results from sampling data uniformly at random from each consistent geometry: as expected SMDS, MDS and HMDS perform dramatically better on data that comes from the geometry it embeds in.}
    \label{fig:sampling_exp}
\end{figure}

In Fig.~\ref{fig:sampling_exp} we  go beyond graphs to verify the different nature of the three geometries. We sample points randomly from each space, and use these points to define the distance matrices. We expect the corresponding geometry's MDS to embed the data with much lower distortion and this is indeed the effect we see.

\section{Conclusions and Future Work}
We described an efficient method for embedding graphs in spherical space. The method generalizes beyond graphs to embedding high-dimensional data.
We studied (quantitatively and qualitatively) the difference between
spherical embeddings of graphs and embeddings in Euclidean and hyperbolic spaces. We discussed the issue of dilation and proposed an approach that seems to work well in practice.
Furthermore, we compared how structures are preserved in different geometries. 
The algorithm is implemented and fully functional and we provide the source code, experimental data and results, and a web based visualization tool on GitHub: \url{https://github.com/Mickey253/spherical-mds}.

%Although spherical geometry offers some benefits for network visualization, its benefits and detriments should be further investigated. For instance, one might think that the finite area on the sphere would mean that spherical node-link diagrams would have particularly poor visual scalability. Additionally, when using the typical orthographic projection a spherical layout may not look and feel all that different from a 3-dimensional Euclidean layout. As such, some of the same problems occur such as half of the sphere being obscured using an Orthographic projection.

% Our investigation in Section~\ref{sec:geo_compare} takes a brute force approach to finding data well suited to the sphere or other geometries. A more formal or systematic method of characterizing graphs that have low distortion in the plane, sphere, or hyperbolic space is an interesting direction for future work. 

While our proposed algorithm is much faster than exact gradient descent (5 seconds for a 1000-vertex graph), it still requires an all-pairs-shortest-paths computation as a preprocessing step, which cannot be done faster than quadratic time in the number of vertices. This is a bottleneck computation for any graph-distance based approach and coming up with a strategy (e.g., sampling a subset of distances) is a problem whose solution can impact many existing algorithms.
Another direction for future work is to quickly determine the best embedding space for a given graph. That is given a graph, decide the best manifold to embed it in: Euclidean, spherical or hyperbolic.
%In Section~\ref{sec:geo_compare} we took a brute force approach to manually compare between the distortion values and picking the geometry which can achieve the smallest distortion for a given graph.
We considered stress and distortion measures here, but exploring other graph drawing aesthetics across different geometries seems to be a worthwhile direction to explore.
%how well the visualization models the underlying data. It is reasonable to think that the sphere may effect the aesthetic criteria of a layout. A user study is a potential avenue of future work.

\typeout{}
\bibliographystyle{splncs04}
\bibliography{smds}

\newpage

\appendix

\noindent \Large{\textbf{Appendix}}

\section{Additional Layouts}
We present additional layouts on the Euclidean plane and on the sphere.

\begin{table*}[htp!] 
  \caption{Layouts} \label{tab:drawings1}
  \centering
  \begin{tabular}
      {l c c |l  c c} \hline & E-MDS & SMDS &  & E-MDS & SMDS  \\
      \hline 
      
      \raisebox{-.5\normalbaselineskip}[0pt][0pt]{\rotatebox[origin=c]{90}{block\_2000}} &
      \parbox[c]{\tablesize\textwidth}{
      \includegraphics[width=\tablesize\textwidth]{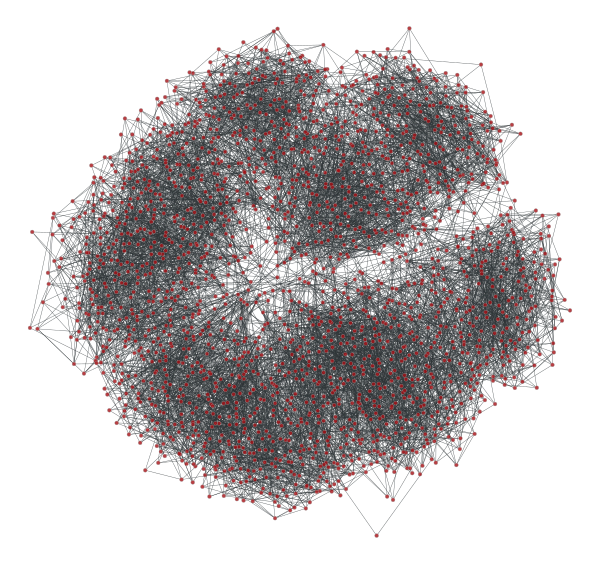}} 
      & \parbox[c]{\tablesize\textwidth}{
      \includegraphics[width=\tablesize\textwidth]{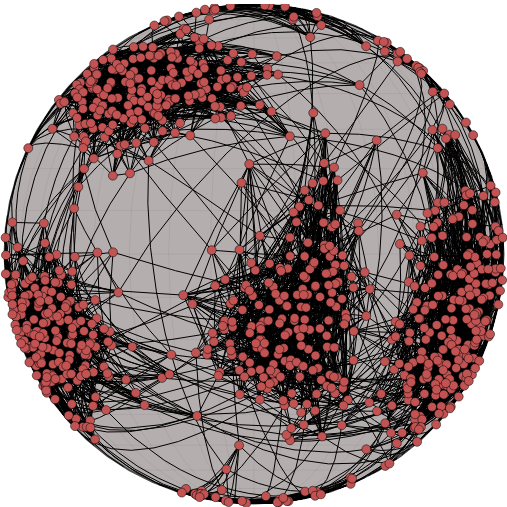}}
      &
      \raisebox{-.5\normalbaselineskip}[0pt][0pt]{\rotatebox[origin=c]{90}{sierpinski3d}} &
      \parbox[c]{\tablesize\textwidth}{
      \includegraphics[width=\tablesize\textwidth]{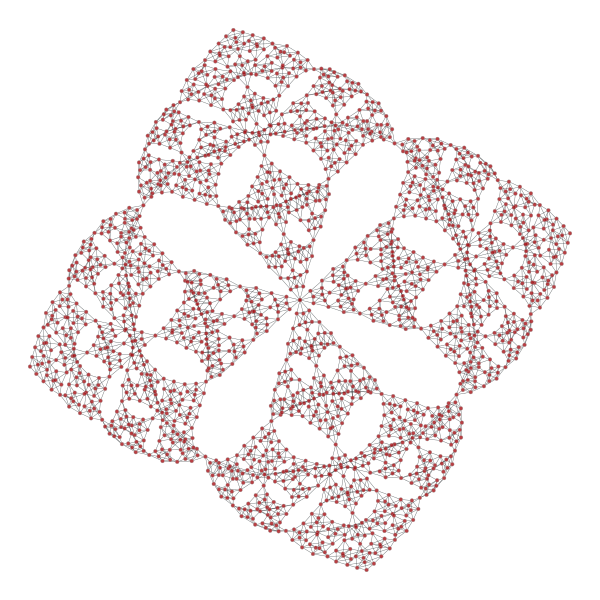}} 
      & \parbox[c]{\tablesize\textwidth}{
      \includegraphics[width=\tablesize\textwidth]{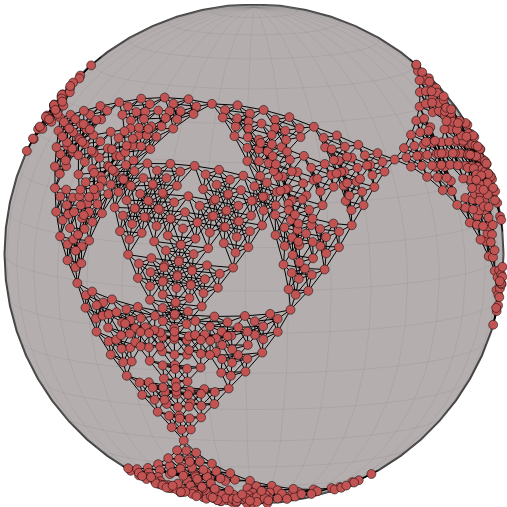}} \\        
      
      \raisebox{-.5\normalbaselineskip}[0pt][0pt]{\rotatebox[origin=c]{90}{CA-GrQc}} &
      \parbox[c]{\tablesize\textwidth}{
      \includegraphics[width=\tablesize\textwidth]{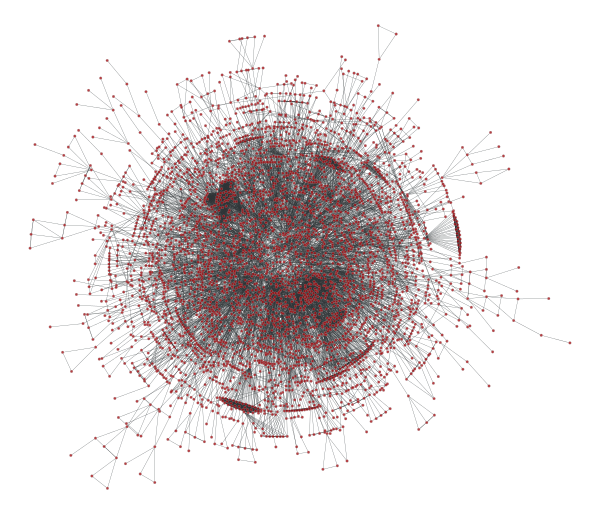}} 
      & \parbox[c]{\tablesize\textwidth}{
      \includegraphics[width=\tablesize\textwidth]{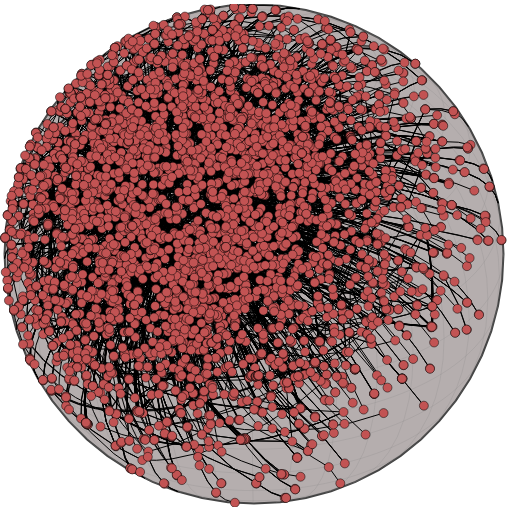}}
      &
      \raisebox{-.5\normalbaselineskip}[0pt][0pt]{\rotatebox[origin=c]{90}{EVA}} &
      \parbox[c]{\tablesize\textwidth}{
      \includegraphics[width=\tablesize\textwidth]{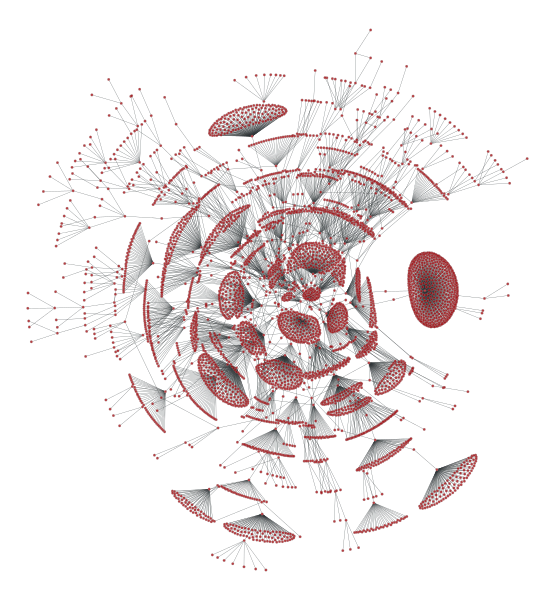}} 
      & \parbox[c]{\tablesize\textwidth}{
      \includegraphics[width=\tablesize\textwidth]{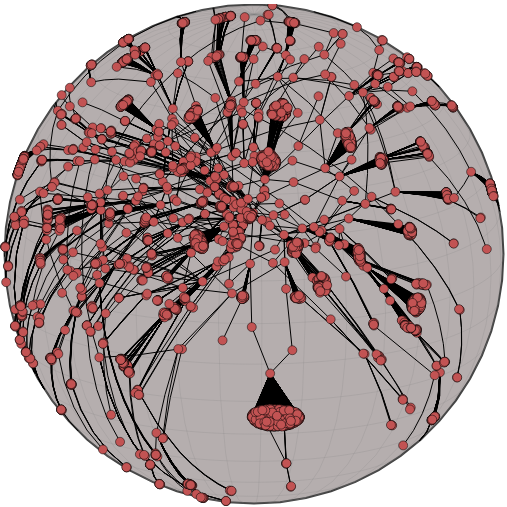}} \\

      \raisebox{-.5\normalbaselineskip}[0pt][0pt]{\rotatebox[origin=c]{90}{3elt}} &
      \parbox[c]{\tablesize\textwidth}{
      \includegraphics[width=\tablesize\textwidth]{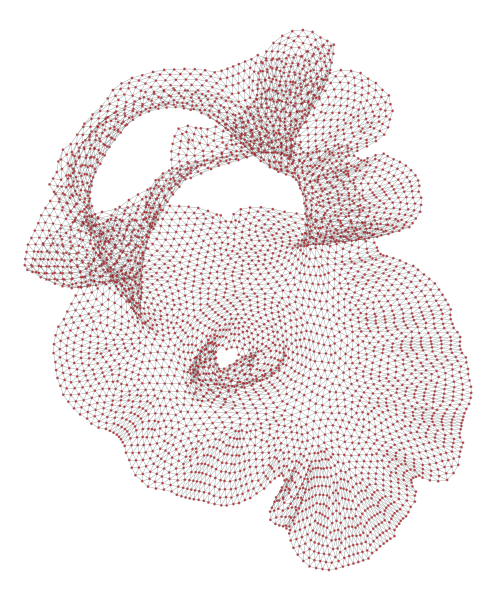}} 
      & \parbox[c]{\tablesize\textwidth}{
      \includegraphics[width=\tablesize\textwidth]{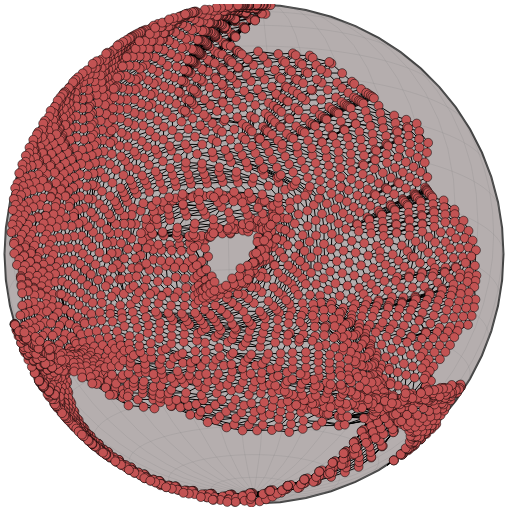}}
      &
      \raisebox{-.5\normalbaselineskip}[0pt][0pt]{\rotatebox[origin=c]{90}{us\_powergrid}} &
      \parbox[c]{\tablesize\textwidth}{
      \includegraphics[width=\tablesize\textwidth]{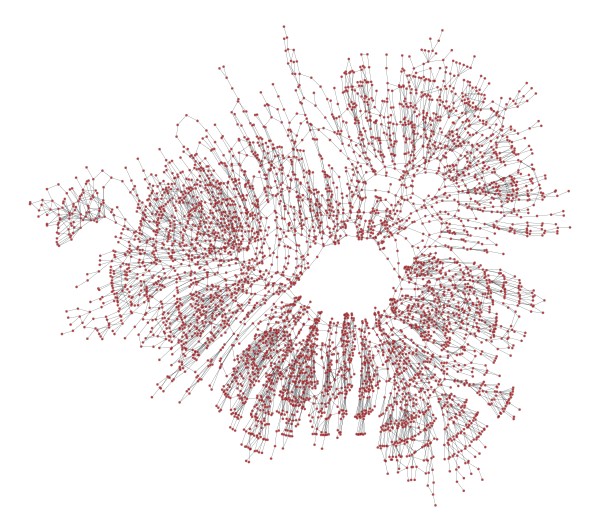}} 
      & \parbox[c]{\tablesize\textwidth}{
      \includegraphics[width=\tablesize\textwidth]{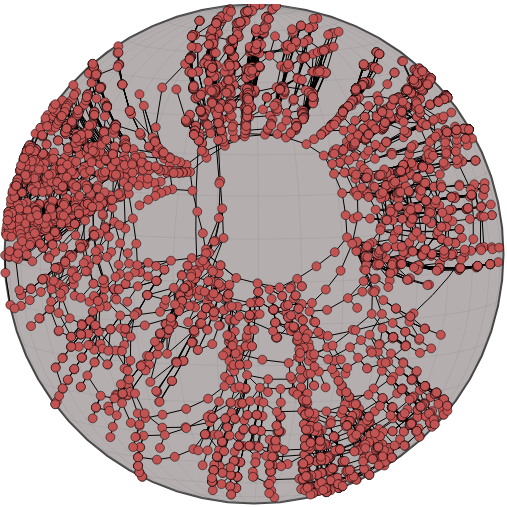}} \\        

      \hline
  \end{tabular}
\end{table*}

\newpage

\begin{table*}[htp!] 
  \caption{Layouts} \label{tab:drawings3}
  \centering
  \begin{tabular}{l c c |l  c c} \hline & E-MDS & SMDS &  & E-MDS & SMDS  \\
      \hline 
      
      \raisebox{-.5\normalbaselineskip}[0pt][0pt]{\rotatebox[origin=c]{90}{cube}} &
      \parbox[c]{\tablesize\textwidth}{
      \includegraphics[width=\tablesize\textwidth]{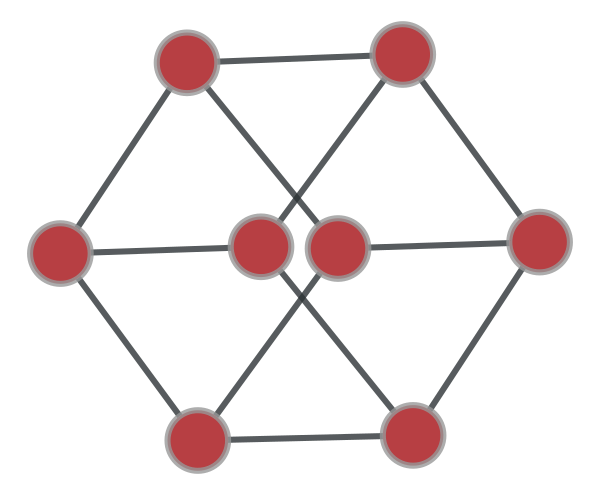}} 
      & \parbox[c]{\tablesize\textwidth}{
      \includegraphics[width=\tablesize\textwidth]{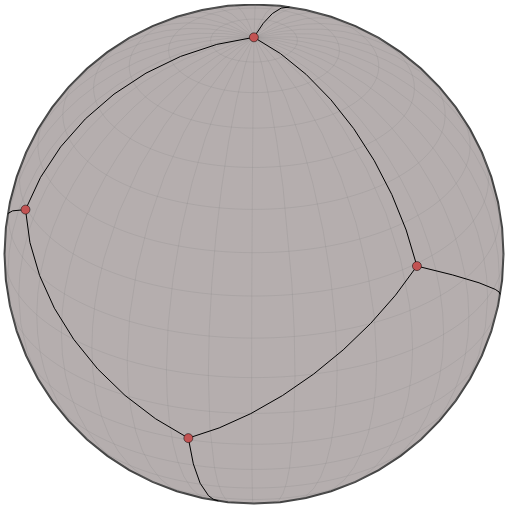}}
      &
      \raisebox{-.5\normalbaselineskip}[0pt][0pt]{\rotatebox[origin=c]{90}{isocahedron}} &
      \parbox[c]{\tablesize\textwidth}{
      \includegraphics[width=\tablesize\textwidth]{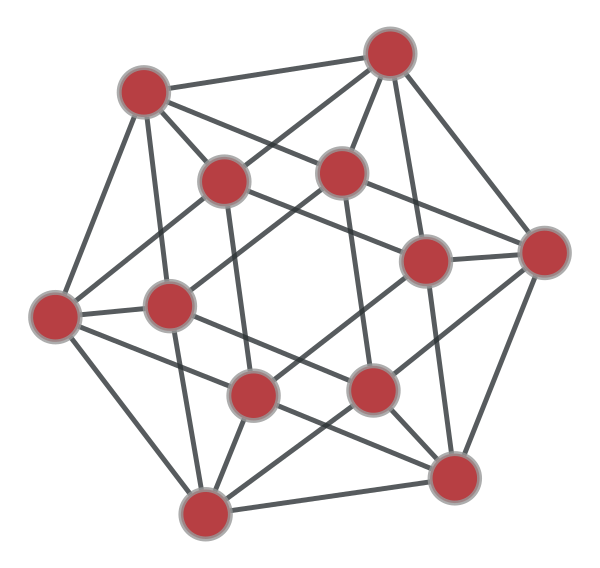}} 
      & \parbox[c]{\tablesize\textwidth}{
      \includegraphics[width=\tablesize\textwidth]{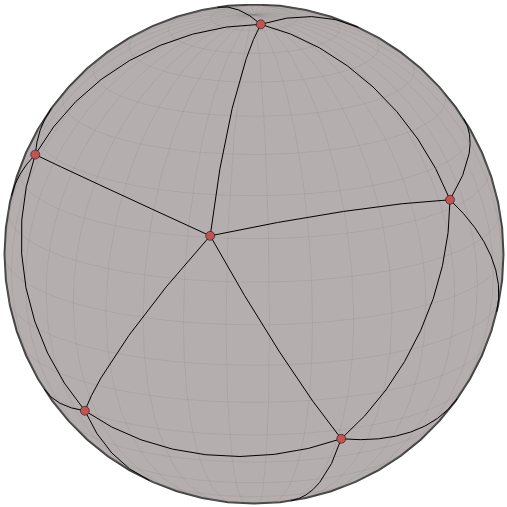}} \\
      
      \raisebox{-.5\normalbaselineskip}[0pt][0pt]{\rotatebox[origin=c]{90}{dodecahedron}} &
      \parbox[c]{\tablesize\textwidth}{
      \includegraphics[width=\tablesize\textwidth]{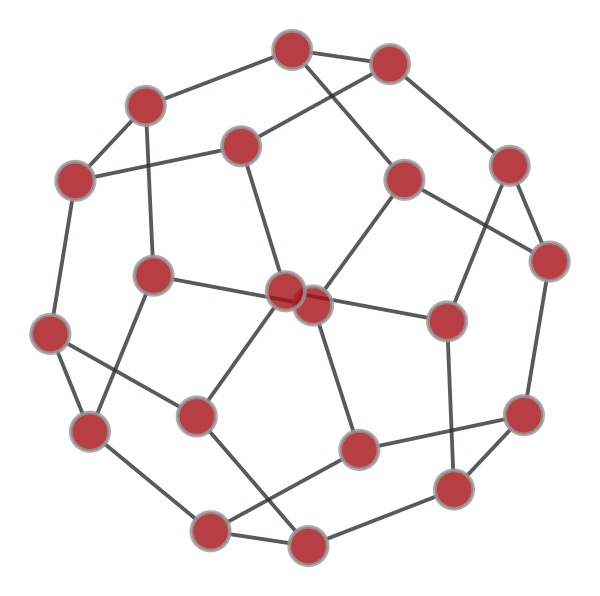}} 
      & \parbox[c]{\tablesize\textwidth}{
      \includegraphics[width=\tablesize\textwidth]{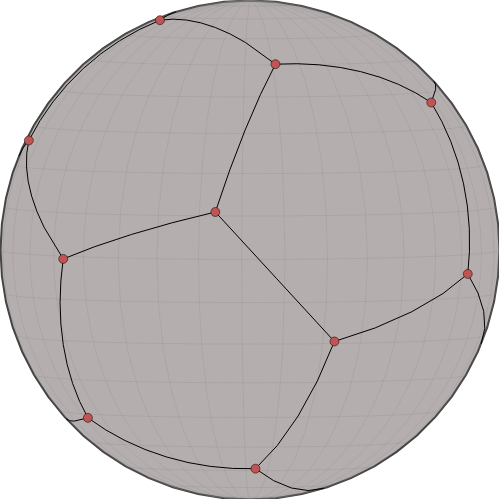}}
    &
      \raisebox{-.5\normalbaselineskip}[0pt][0pt]{\rotatebox[origin=c]{90}{dwt\_72}} &
      \parbox[c]{\tablesize\textwidth}{
      \includegraphics[width=\tablesize\textwidth]{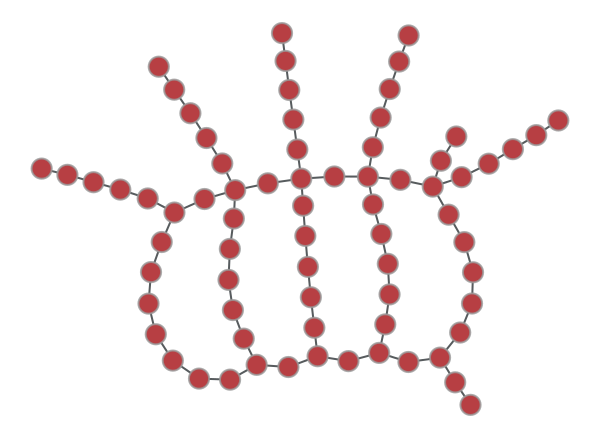}} 
      & \parbox[c]{\tablesize\textwidth}{
      \includegraphics[width=\tablesize\textwidth]{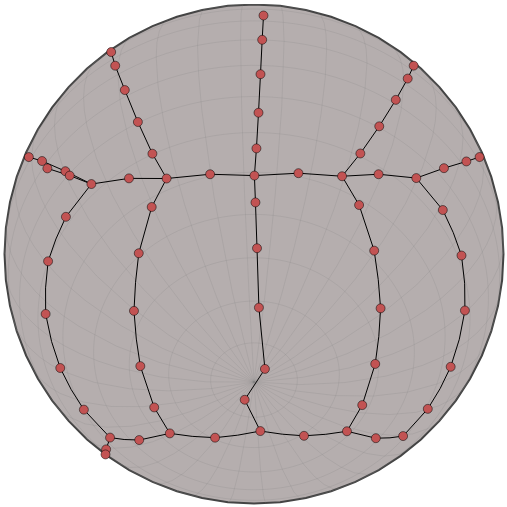}} \\  
      
      \raisebox{-.5\normalbaselineskip}[0pt][0pt]{\rotatebox[origin=c]{90}{lesmis}} &
      \parbox[c]{\tablesize\textwidth}{
      \includegraphics[width=\tablesize\textwidth]{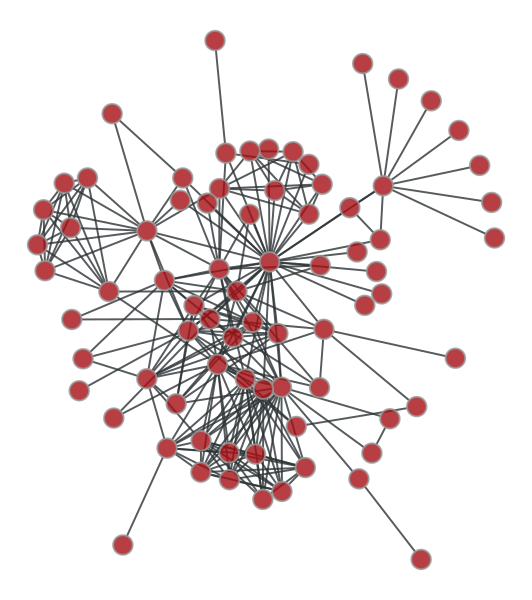}} 
      & \parbox[c]{\tablesize\textwidth}{
      \includegraphics[width=\tablesize\textwidth]{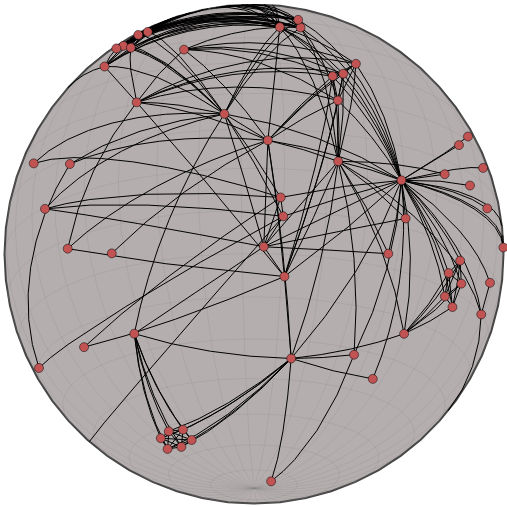}}
    &
      \raisebox{-.5\normalbaselineskip}[0pt][0pt]{\rotatebox[origin=c]{90}{jazz}} &
      \parbox[c]{\tablesize\textwidth}{
      \includegraphics[width=\tablesize\textwidth]{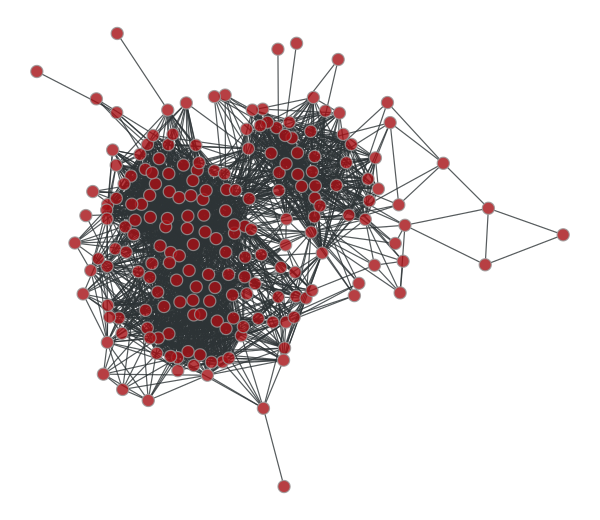}} 
      & \parbox[c]{\tablesize\textwidth}{
      \includegraphics[width=\tablesize\textwidth]{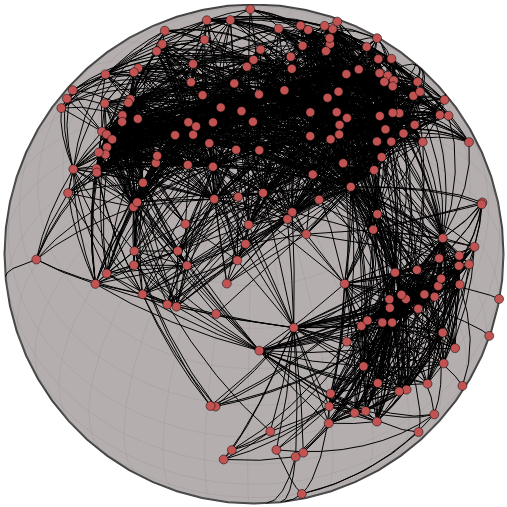}} \\ 
      
      \raisebox{-.5\normalbaselineskip}[0pt][0pt]{\rotatebox[origin=c]{90}{visbrazil}} &
      \parbox[c]{\tablesize\textwidth}{
      \includegraphics[width=\tablesize\textwidth]{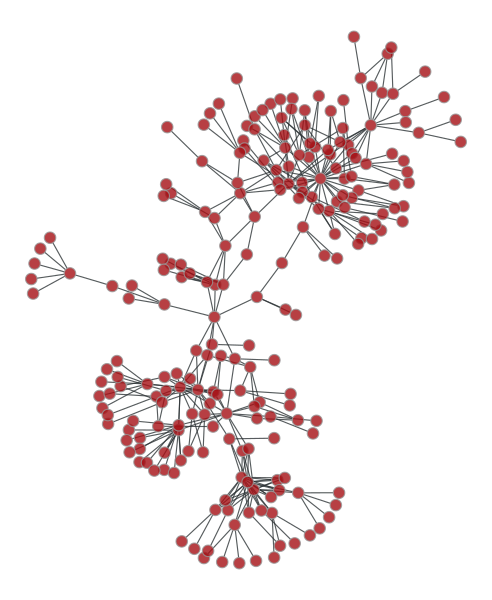}} 
      & \parbox[c]{\tablesize\textwidth}{
      \includegraphics[width=\tablesize\textwidth]{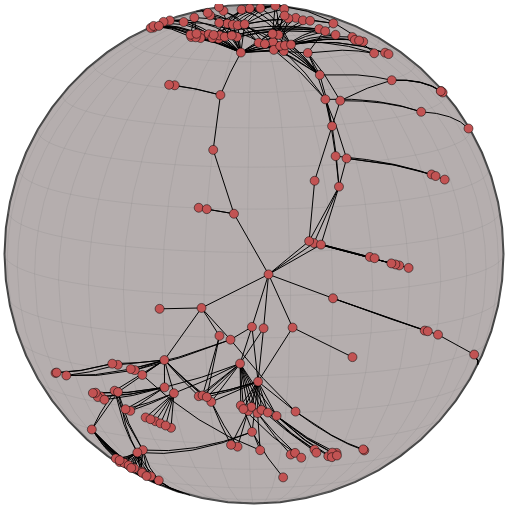}}
    &
      \raisebox{-.5\normalbaselineskip}[0pt][0pt]{\rotatebox[origin=c]{90}{grid1\_dual}} &
      \parbox[c]{\tablesize\textwidth}{
      \includegraphics[width=\tablesize\textwidth]{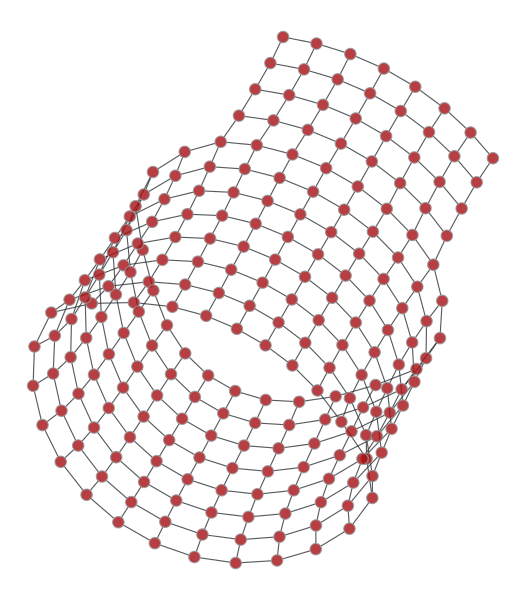}} 
      & \parbox[c]{\tablesize\textwidth}{
      \includegraphics[width=\tablesize\textwidth]{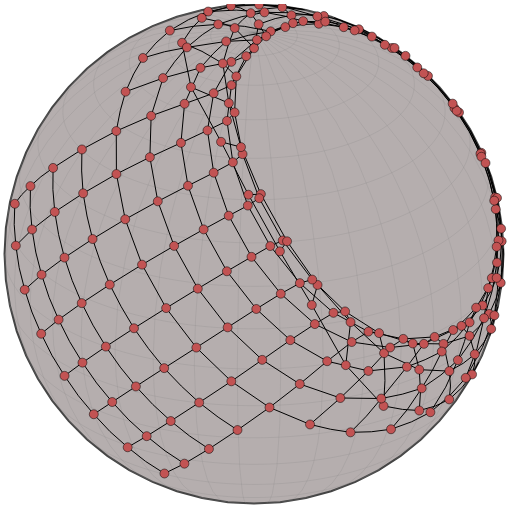}} \\ 
      
      \raisebox{-.5\normalbaselineskip}[0pt][0pt]{\rotatebox[origin=c]{90}{dwt\_307}} &
      \parbox[c]{\tablesize\textwidth}{
      \includegraphics[width=\tablesize\textwidth]{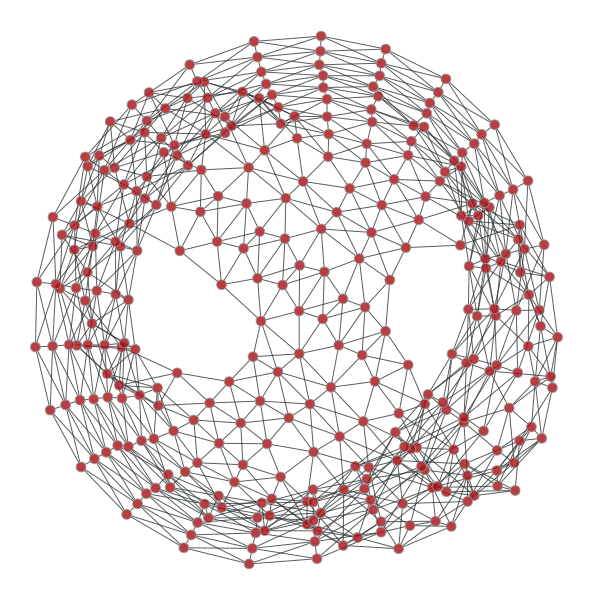}} 
      & \parbox[c]{\tablesize\textwidth}{
      \includegraphics[width=\tablesize\textwidth]{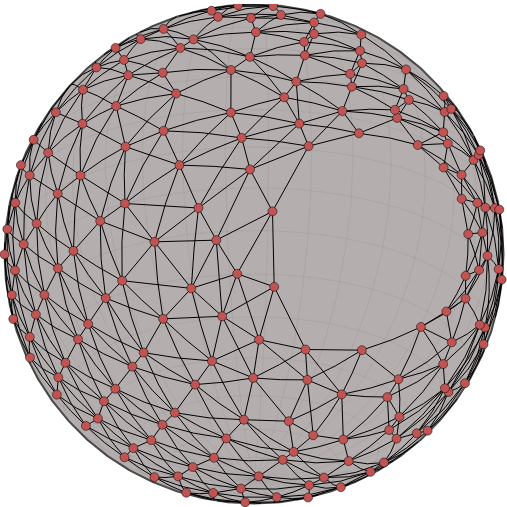}}
    &
      \raisebox{-.5\normalbaselineskip}[0pt][0pt]{\rotatebox[origin=c]{90}{dwt\_361}} &
      \parbox[c]{\tablesize\textwidth}{
      \includegraphics[width=\tablesize\textwidth]{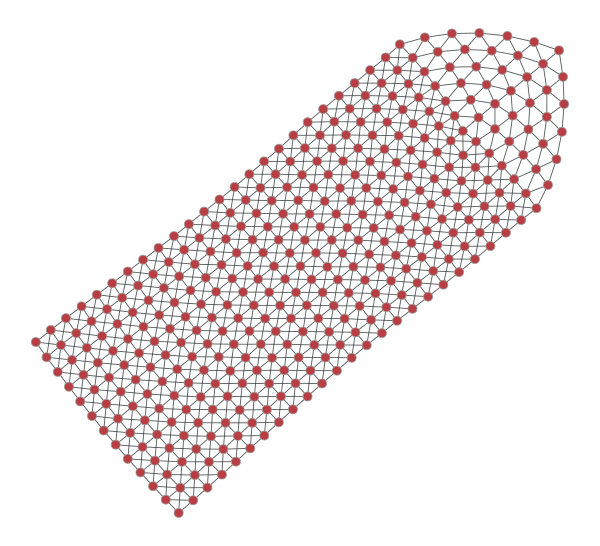}} 
      & \parbox[c]{\tablesize\textwidth}{
      \includegraphics[width=\tablesize\textwidth]{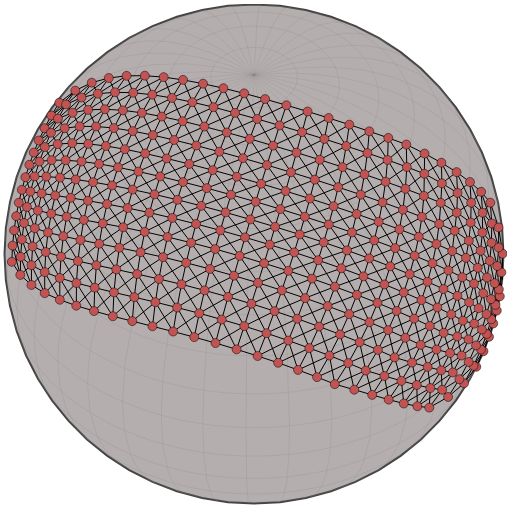}} \\
      
      \raisebox{-.5\normalbaselineskip}[0pt][0pt]{\rotatebox[origin=c]{90}{netscience}} &
      \parbox[c]{\tablesize\textwidth}{
      \includegraphics[width=\tablesize\textwidth]{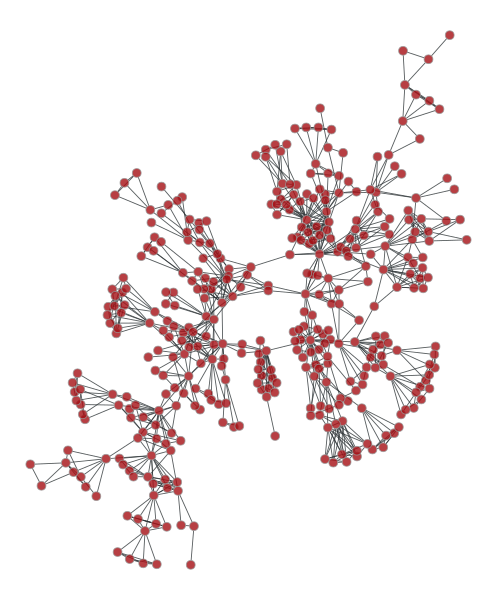}} 
      & \parbox[c]{\tablesize\textwidth}{
      \includegraphics[width=\tablesize\textwidth]{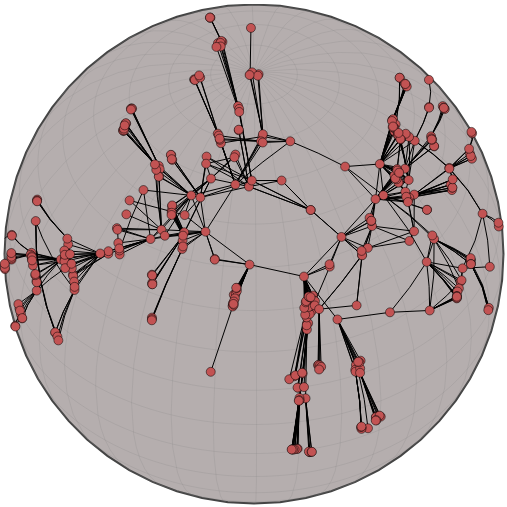}}
    &
      \raisebox{-.5\normalbaselineskip}[0pt][0pt]{\rotatebox[origin=c]{90}{dwt\_419}} &
      \parbox[c]{\tablesize\textwidth}{
      \includegraphics[width=\tablesize\textwidth]{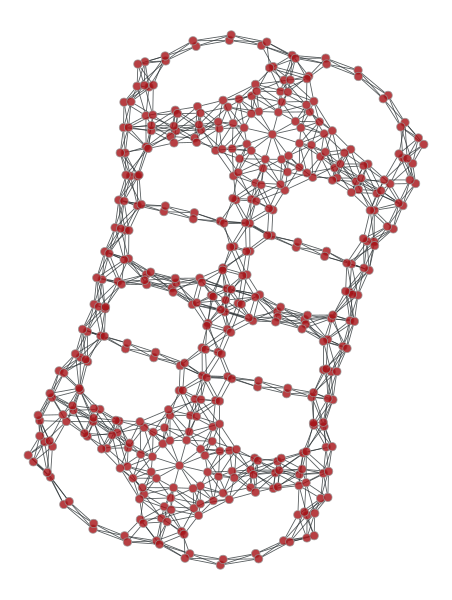}} 
      & \parbox[c]{\tablesize\textwidth}{
      \includegraphics[width=\tablesize\textwidth]{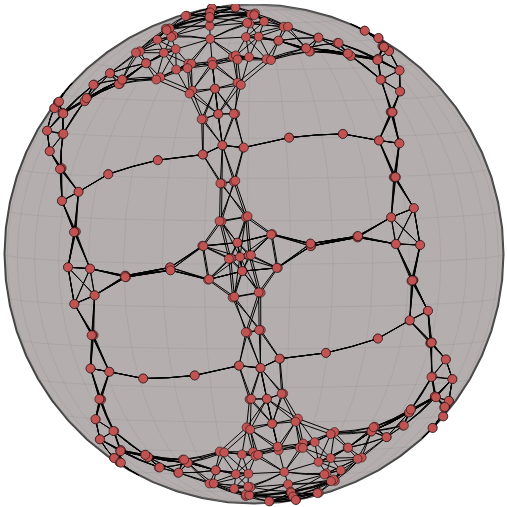}} \\  
      
      \raisebox{-.5\normalbaselineskip}[0pt][0pt]{\rotatebox[origin=c]{90}{oscil\_dcop\_01}} &
      \parbox[c]{\tablesize\textwidth}{
      \includegraphics[width=\tablesize\textwidth]{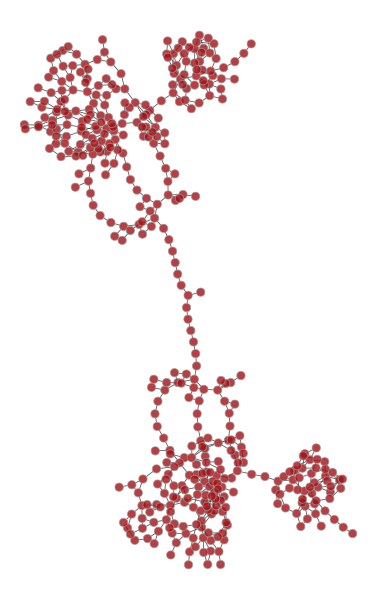}} 
      & \parbox[c]{\tablesize\textwidth}{
      \includegraphics[width=\tablesize\textwidth]{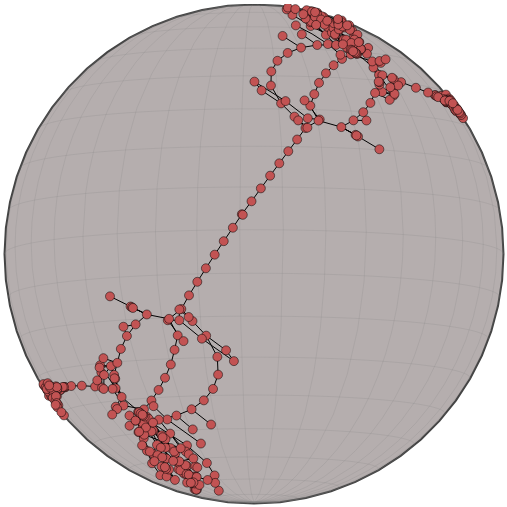}}
    &
      \raisebox{-.5\normalbaselineskip}[0pt][0pt]{\rotatebox[origin=c]{90}{can\_445}} &
      \parbox[c]{\tablesize\textwidth}{
      \includegraphics[width=\tablesize\textwidth]{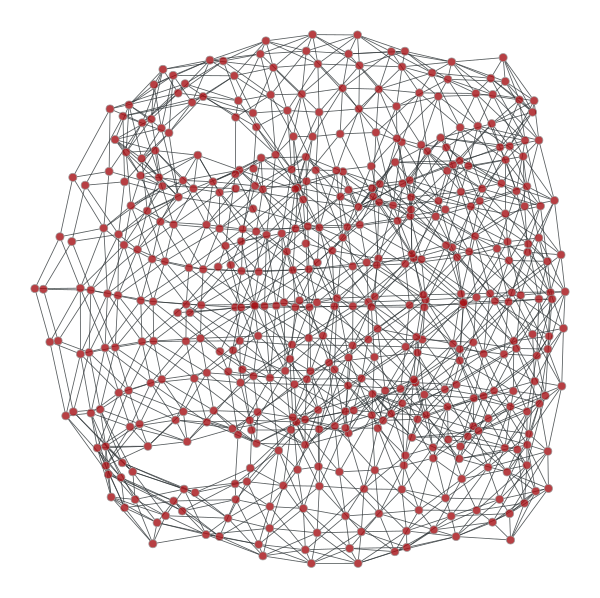}} 
      & \parbox[c]{\tablesize\textwidth}{
      \includegraphics[width=\tablesize\textwidth]{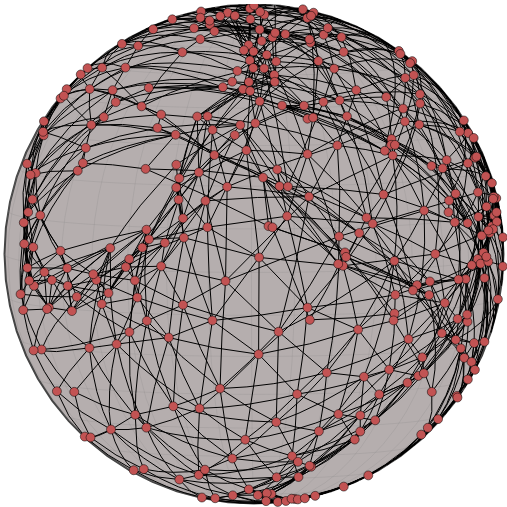}} \\

      \hline
  \end{tabular}
\end{table*}

\newpage

\begin{table*}[htp!] 
  \caption{Layouts} \label{tab:drawings2}
  \centering
  \begin{tabular}
      {l c c |l  c c} \hline & E-MDS & SMDS &  & E-MDS & SMDS  \\
      \hline 
      
      \raisebox{-.5\normalbaselineskip}[0pt][0pt]{\rotatebox[origin=c]{90}{494\_bus}} &
      \parbox[c]{\tablesize\textwidth}{
      \includegraphics[width=\tablesize\textwidth]{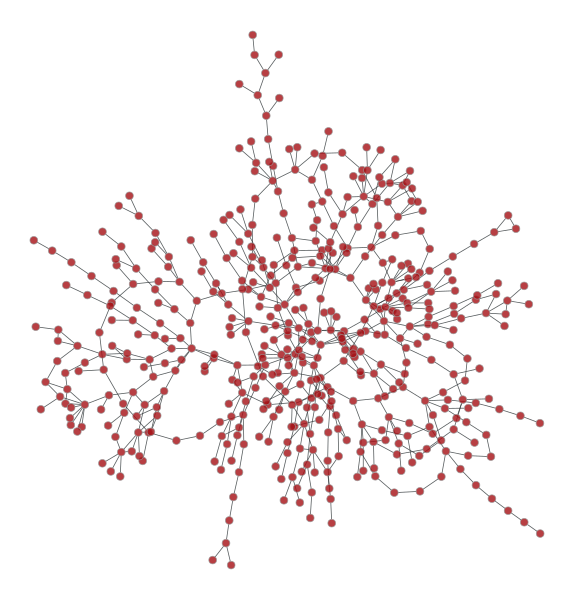}} 
      & \parbox[c]{\tablesize\textwidth}{
      \includegraphics[width=\tablesize\textwidth]{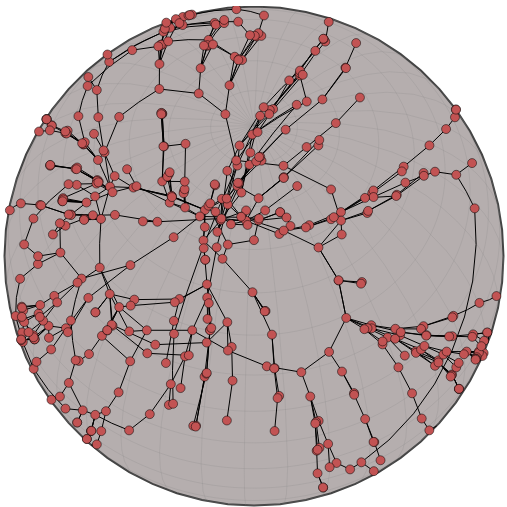}}
      &
      \raisebox{-.5\normalbaselineskip}[0pt][0pt]{\rotatebox[origin=c]{90}{dwt\_918}} &
      \parbox[c]{\tablesize\textwidth}{
      \includegraphics[width=\tablesize\textwidth]{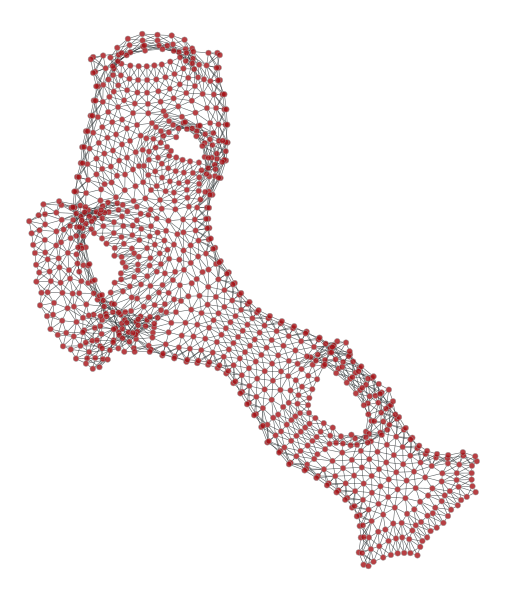}} 
      & \parbox[c]{\tablesize\textwidth}{
      \includegraphics[width=\tablesize\textwidth]{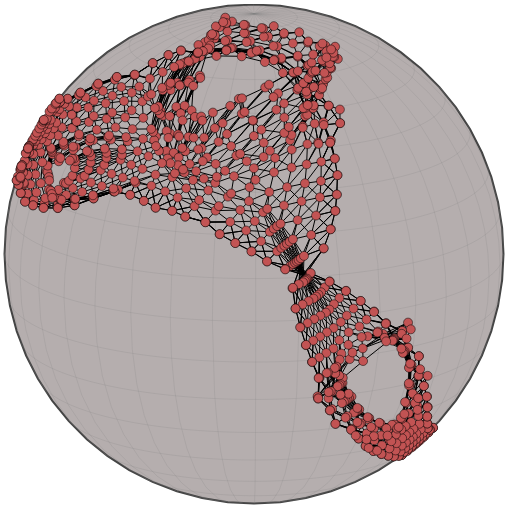}} \\
      
      \raisebox{-.5\normalbaselineskip}[0pt][0pt]{\rotatebox[origin=c]{90}{price\_1000}} &
      \parbox[c]{\tablesize\textwidth}{
      \includegraphics[width=\tablesize\textwidth]{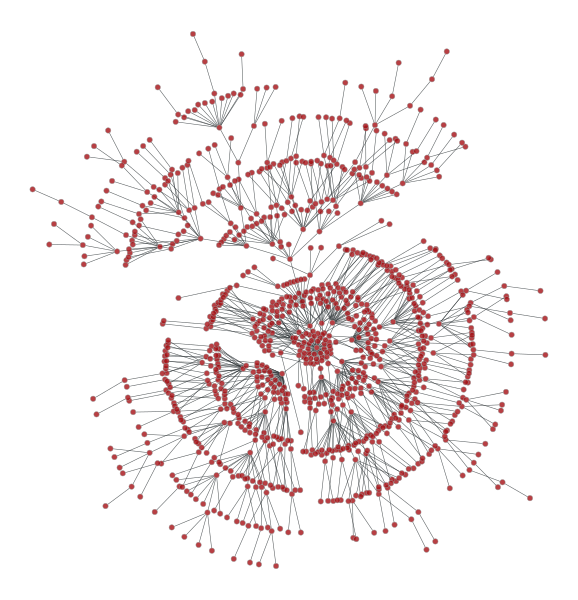}} 
      & \parbox[c]{\tablesize\textwidth}{
      \includegraphics[width=\tablesize\textwidth]{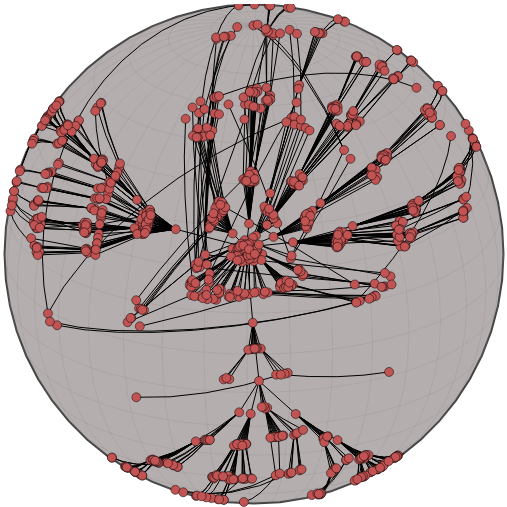}}
      &
      \raisebox{-.5\normalbaselineskip}[0pt][0pt]{\rotatebox[origin=c]{90}{dwt\_1005}} &
      \parbox[c]{\tablesize\textwidth}{
      \includegraphics[width=\tablesize\textwidth]{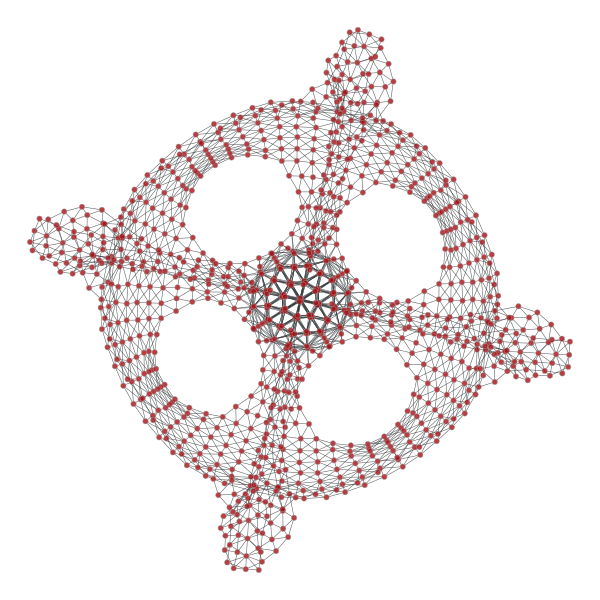}} 
      & \parbox[c]{\tablesize\textwidth}{
      \includegraphics[width=\tablesize\textwidth]{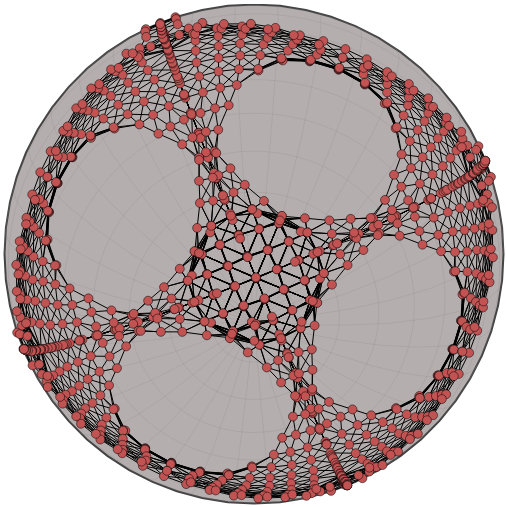}} \\
      
      \raisebox{-.5\normalbaselineskip}[0pt][0pt]{\rotatebox[origin=c]{90}{cage8}} &
      \parbox[c]{\tablesize\textwidth}{
      \includegraphics[width=\tablesize\textwidth]{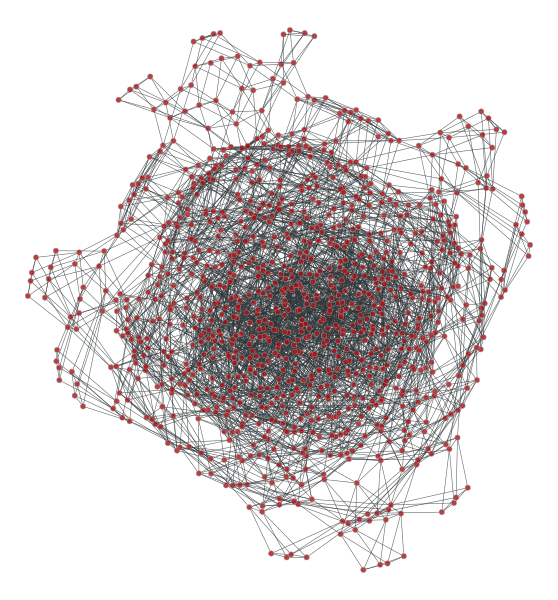}} 
      & \parbox[c]{\tablesize\textwidth}{
      \includegraphics[width=\tablesize\textwidth]{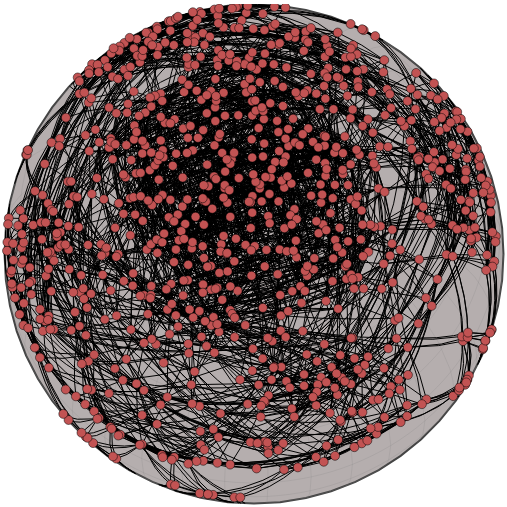}}
      &
      \raisebox{-.5\normalbaselineskip}[0pt][0pt]{\rotatebox[origin=c]{90}{delaunay\_n10}} &
      \parbox[c]{\tablesize\textwidth}{
      \includegraphics[width=\tablesize\textwidth]{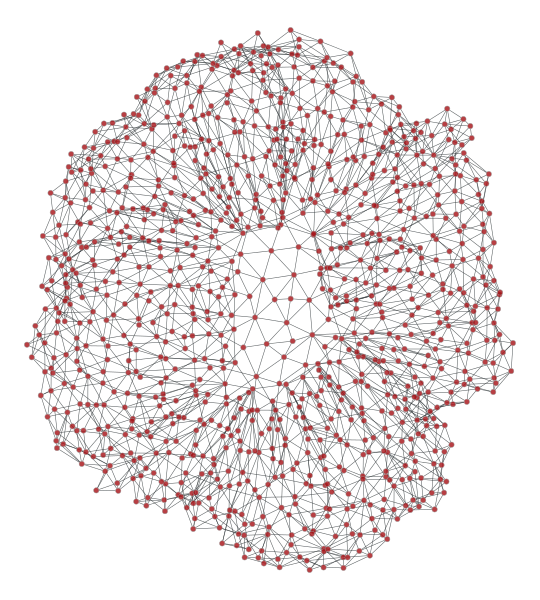}} 
      & \parbox[c]{\tablesize\textwidth}{
      \includegraphics[width=\tablesize\textwidth]{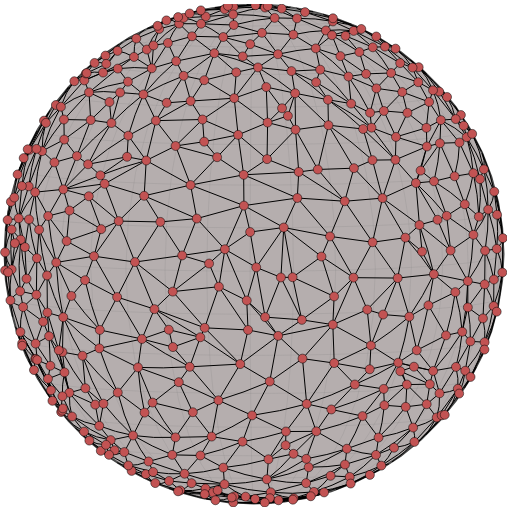}} \\

      \raisebox{-.5\normalbaselineskip}[0pt][0pt]{\rotatebox[origin=c]{90}{can\_144}} &
      \parbox[c]{\tablesize\textwidth}{
      \includegraphics[width=\tablesize\textwidth]{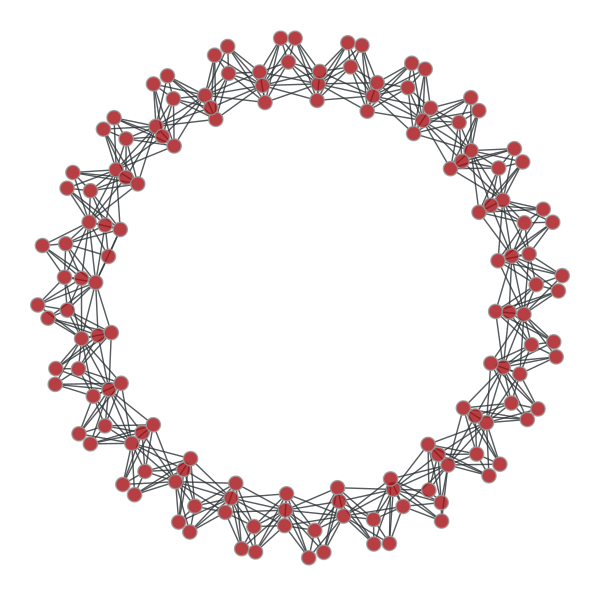}} 
      & \parbox[c]{\tablesize\textwidth}{
      \includegraphics[width=\tablesize\textwidth]{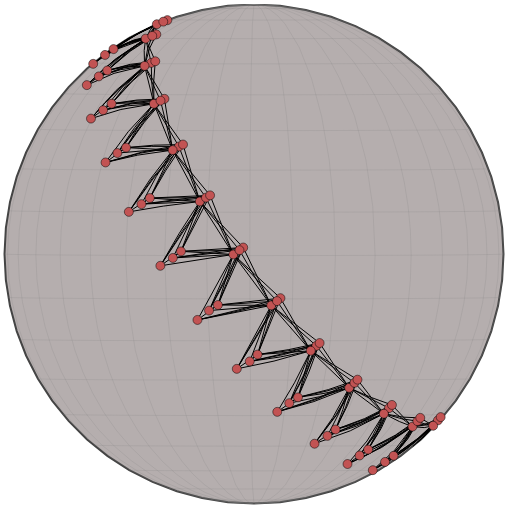}}
      &
      \raisebox{-.5\normalbaselineskip}[0pt][0pt]{\rotatebox[origin=c]{90}{fpga}} &
      \parbox[c]{\tablesize\textwidth}{
      \includegraphics[width=\tablesize\textwidth]{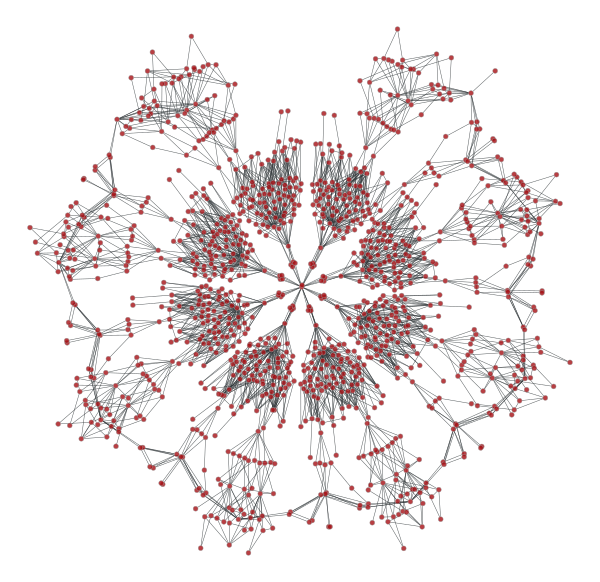}} 
      & \parbox[c]{\tablesize\textwidth}{
      \includegraphics[width=\tablesize\textwidth]{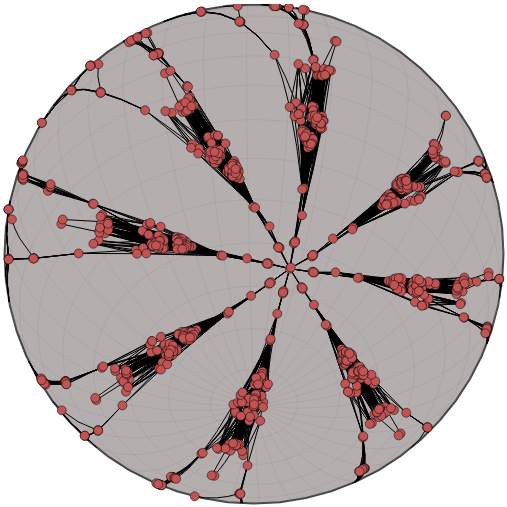}} \\

      \raisebox{-.5\normalbaselineskip}[0pt][0pt]{\rotatebox[origin=c]{90}{dwt\_162}} &
      \parbox[c]{\tablesize\textwidth}{
      \includegraphics[width=\tablesize\textwidth]{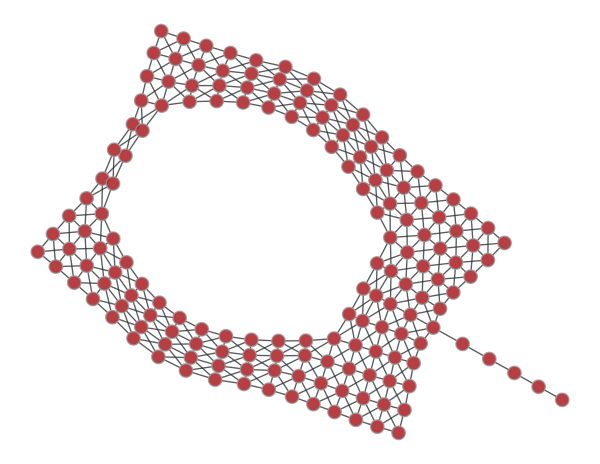}} 
      & \parbox[c]{\tablesize\textwidth}{
      \includegraphics[width=\tablesize\textwidth]{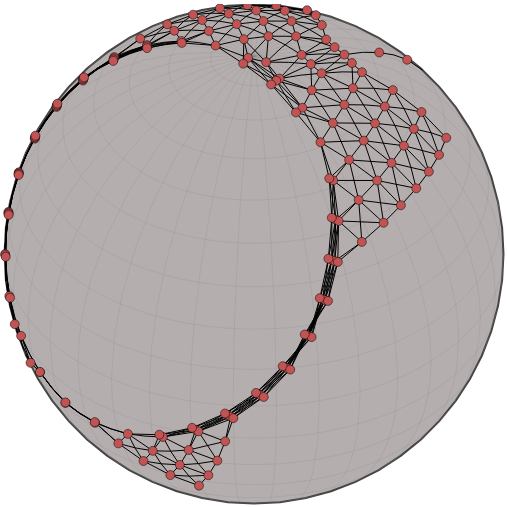}}      
      &
      \raisebox{-.5\normalbaselineskip}[0pt][0pt]{\rotatebox[origin=c]{90}{dodecahedron\_4}} &
      \parbox[c]{\tablesize\textwidth}{
      \includegraphics[width=\tablesize\textwidth]{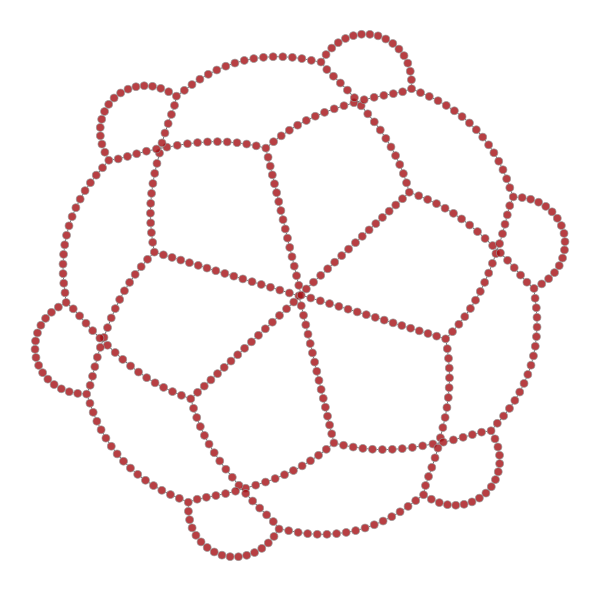}} 
      & \parbox[c]{\tablesize\textwidth}{
      \includegraphics[width=\tablesize\textwidth]{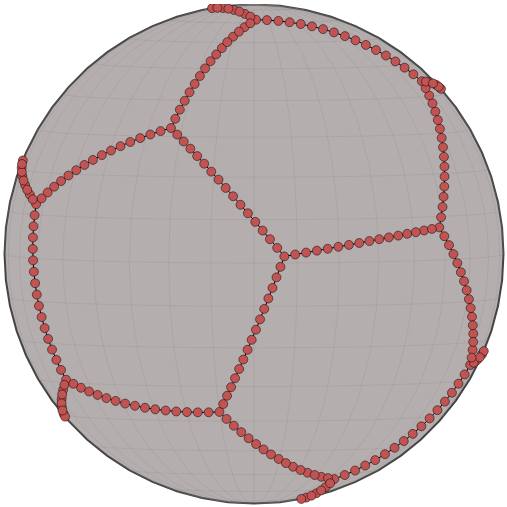}} \\        
      
      \raisebox{-.5\normalbaselineskip}[0pt][0pt]{\rotatebox[origin=c]{90}{isocahedron\_4}} &
      \parbox[c]{\tablesize\textwidth}{
      \includegraphics[width=\tablesize\textwidth]{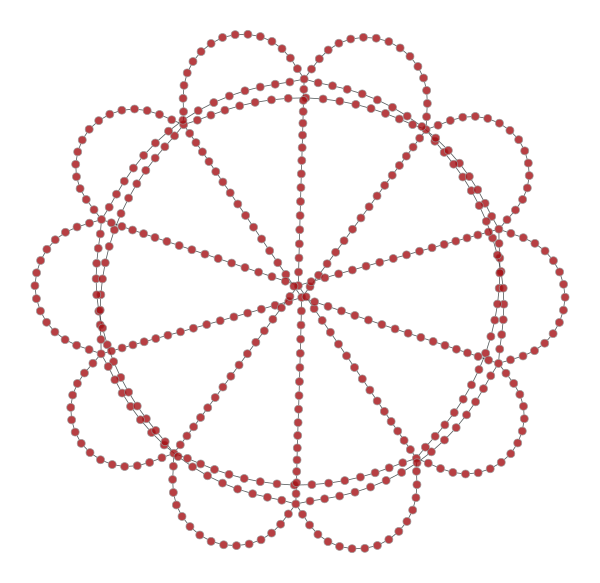}} 
      & \parbox[c]{\tablesize\textwidth}{
      \includegraphics[width=\tablesize\textwidth]{figures/projections/iso_ortho.png}}
      &
      &
    
      & \\              

      \hline
  \end{tabular}
  
\end{table*}

\section{Further geometry comparison}
\label{sec:appendix-geo-compare}

\begin{figure}[htp!]
    \centering 
    \includegraphics[width=0.6\textwidth]{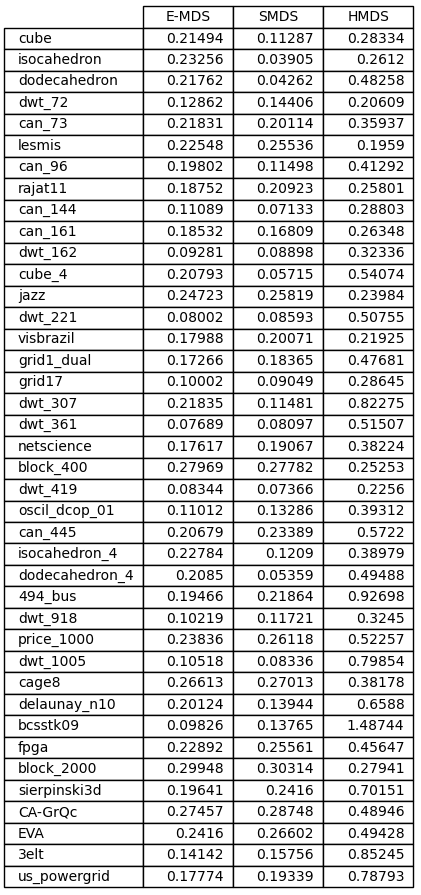}
    \caption{Average distortion values on each graph for each geometry.}
    \label{fig:distortion-table}
\end{figure}

\end{document}